# Virtual Relationships:
# Short- and Long-run Evidence from BitCoin and Altcoin Markets[1]

*Pavel Ciaian*[♠,↗], *Miroslava Rajcaniova*[♯,𝛾], *d'Artis Kancs*[♠,↗]

[♠]European Commission (DG JRC)
[↗]Catholic University of Leuven (LICOS)
[♯]Slovak University of Agriculture (SUA)
[𝛾]University of West Bohemia (UWB)

**Abstract:** This study empirically examines interdependencies between BitCoin and altcoin markets in the short- and long-run. We apply time-series analytical mechanisms to daily data of 17 virtual currencies (BitCoin + 16 alternative virtual currencies) and two Altcoin price indices for the period 2013-2016. Our empirical findings confirm that indeed BitCoin and Altcoin markets are interdependent. The BitCoin-Altcoin price relationship is significantly stronger in the short-run than in the long-run. We cannot fully confirm the hypothesis that the BitCoin price relationship is stronger with those Altcoins that are more similar in their price formation mechanism to BitCoin. In the long-run, macro-financial indicators determine the altcoin price formation to a greater degree than BitCoin does. The virtual currency supply is exogenous and therefore plays only a limited role in the price formation.

[1] We gratefully acknowledge financial support received from the Slovak Research and Development Agency under the contract No. APVV-15-0552 and VEGA 1/0797/16. The authors are solely responsible for the content of the paper. The views expressed are purely those of the authors and may not in any circumstances be regarded as stating an official position of the European Commission.

# 1 Introduction

Since around a decade, virtual currencies represent a new phenomenon on global financial markets. By providing an alternative money and investment opportunity, they function outside centralized financial institutions. While offering a less expensive alternative to mainstream currencies in terms of transaction costs, the prices of virtual currencies are developing considerably more erratically and fluctuate much wider than those of standard currencies (Bouoiyour, Selmi and Tiwari 2014; Ciaian, Rajcaniova and Kancs 2016b). The objective of the present study is to analyze the interdependencies between different virtual currency prices and test short- and long-run relationships between BitCoin and most important BitCoin alternatives.

BitCoin was the first decentralized ledger currency; it was created in 2009. Since its introduction, BitCoin continues to be the most widely used virtual currency and is the largest in terms of the market value, the total virtual currency market capitalization and the number of daily transactions. Also the number of businesses accepting BitCoin is increasing continuously. The rise of BitCoin has triggered large interest not only among virtual currency users and investors, but also in the scientific literature that has analyzed extensively the price formation of BitCoin (e.g. Grinberg, 2011; Barber et al., 2012; Kroll, Davey and Felten, 2013; Moore and Christin, 2013; Bouoiyour, Selmi and Tiwari, 2014; Kristoufek, 2015; Ciaian, Rajcaniova and Kancs, 2016 a). These studies have identified a number of determinants of the BitCoin price development in the long-run: (i) market forces of the BitCoin supply and demand (Buchholz et al., 2012; Bouoiyour and Selmi, 2015); (ii) the BitCoin's attractiveness for investors (Kristoufek, 2013; Bouoiyour and Selmi, 2015); and (iii) the influence of global macro-financial developments (van Wijk, 2013).

The success of BitCoin has led to the emergence of many alternative virtual currencies (altcoins),[2] such as LiteCoin, PeerCoin, AuroraCoin, DogeCoin, Ripple, etc. Most of altcoins rely on the same or similar blockchain technology as BitCoin, and aim to either complement or improve certain BitCoin characteristics. For example, Litecoin aims to save the computing power required for the coin mining; Peercoin aims to improve the efficiency of mining and the currency's security; Dash aims at a faster processing of transactions and offers an enhanced privacy protection; Bitshares and Ethereum provide additional features, such as a digital platform to run smart contracts. Being new, decentralized and small virtual currencies without a monetary base, altcoin prices fluctuate considerably wider than those of standard fiat currencies, such as the United States Dollar, Euro and many others (IMF, 2016). Despite the comparably high market volatility, there is little known about their price formation mechanisms and altcoin interdependencies with the BitCoin market. Indeed, there are good reasons to believe that BitCoin and altcoin prices might be interdependent, given that BitCoin is the dominant virtual currency, similar patterns in BitCoin and altcoin price developments, as well as the fact that a large majority of altcoin purchases are executed in BitCoins.

The present study attempts to fill this knowledge gap about the altcoin price formation by studying price interdependencies between BitCoin and altcoin markets in the short- and long-run. We analyze prices of 6 major altcoins, 10 minor altcoins and two altcoin price indices. In particular, we test two hypotheses related to BitCoin and altcoin market interdependencies.

---

[2] Alternative virtual currencies are often referred to as "altcoins", which is an abbreviation of "BitCoin alternative".

Hypothesis 1 says that the prices of Altcoins are driven by the BitCoin price development. Hypothesis 2 says that similarities in the BitCoin and Altcoin price formation mechanisms strengthen cointegration between markets. Following the previous literature on virtual currencies, we also control for global macro-financial developments in determining the prices of virtual currencies (van Wijk, 2013; Ciaian, Rajcaniova and Kancs, 2016 a).

Our empirical findings for Hypothesis 1 suggest that BitCoin and altcoin prices are indeed interdependent, though more in the short- than in the long-run. In the short-run, 15 out of 16 studied altcoin prices are impacted by real shocks to BitCoin prices. In the long-run, however, only 4 altcoins are cointegrated with BitCoin prices. Thus, we cannot reject Hypothesis 1 for the short-run, while we reject Hypothesis 1 for the long-run BitCoin-Altcoin relationship for most altcoins.

The empirical results for Hypothesis 2 are more nuanced. We find a weak support for the hypothesis that more similar altcoins to BitCoin respond stronger to real shocks to the BitCoin price in the short-run compared to dissimilar altcoins. However, given that BitCoin consistently impacts all altcoins in the short-run, our results suggest that Hypothesis 1 tends to dominate over Hypothesis 2. In the long-run, however, the estimation results show that differences/similarities of altcoins with BitCoin cannot explain the long-run relationship between BitCoin and altcoin prices. Thus, we have to reject Hypothesis 2 for the long-run BitCoin-Altcoin relationship.

As regards the virtual currency supply and demand, their impact on virtual currency prices (both BitCoin and altcoins) is stronger in the short- than in the long-run. Similarly, the impact of global macroeconomic and financial developments on virtual currency prices has a somehow stronger impact in the short- than the long-run. Particularly, the gold price, exchange rates (USD/EUR and CNY/USD) and the 10-Year Treasury Constant Maturity Rate have a statistically significant impact on BitCoin and altcoin prices in the short- and long-run.

This study is the first paper in the literature analyzing price interdependencies between BitCoin and Altcoin markets in the short- and long-run. This is our main contribution and value added to the existing literature on virtual currencies.

The rest of the paper is structured as follows. In section 2 we derive testable hypotheses about interdependencies of BitCoin and altcoin markets. Section 3 details data sources and the variable construction. The estimation methodology is outlined in Section 4. Our estimation results are presented in section 5, whereas the final section concludes.

## 2 Testable hypotheses

### *Hypothesis 1: The prices of Altcoins are driven by the BitCoin price development*

The central question that we aim to answer in this study is whether BitCoin drives the prices of altcoins. The BitCoin price development can be expected to impact altcoin prices for at least two reasons. First, because BitCoin dominates (*market size*) the virtual currency market with around 80% of the market share (Figure 2), compared to altcoins, in many occasions it can be expected to be the preferred medium of the exchange of goods and services for businesses and consumers. Similarly, because of its dominant market position, BitCoin might often be also the preferred investment asset for potential investors. Both trade-offs between BitCoin and altcoins may cause price interdependencies. Second, the strong dominance of BitCoin as a medium of exchange in altcoin sales transactions (*trading currency composition*) suggests a further source of interdependencies between BitCoin and altcoin prices. Third, BitCoin and altcoin markets show similar price developments (*similarities in price developments*). This suggests that either BitCoin is driving altcoin prices or both prices are confounded e.g. by macro-financial developments.

*Market size*

The market size of virtual currencies has shown a strong growth over last years. The total market capitalization represented around $12.6 billion at the end of 2016, increasing by around eight-fold from the begging of 2013 (from $1.5 billion) (Figure 1). The main driver behind this rapid growth is BitCoin, as the BitCoin market share represents around 80%. Although, the market size of altcoins has increased significantly during recent years, both in relative and absolute terms, altcoins have a rather minor share in the total market capitalization of virtual currencies, their importance continues increasing steadily in recent years: from 5.5% of the total market capitalization at the begging of 2013 to around 20% at the end of 2016. Ethereum is the largest altcoin with a market share of around 8%. Other altcoins have a market share smaller than 3% (Figure 2). Given that BitCoin is the best known and most widely used virtual currency, compared to altcoins, it can be expected to be the preferred medium of exchange of goods and services (i.e. to serve as a money) for businesses and consumers. Altcoins, which represent alternative virtual currencies to BitCoin, compete as a medium of exchange in the same market segment for being accepted as a universal currency. Trade-offs between the use of BitCoin versus altcoins may lead to interdependencies between BitCoin and altcoin markets. Similarly, virtual currencies are also considered as investment assets, whereby potential investors weigh expected benefits from investing in altcoins relative to BitCoin. Because of its large market share, BitCoin is being traded off alternative virtual currencies, which may also cause price interdependencies between altcoins and BitCoin.

*Trading currency composition*

Regarding the regional distribution and the trading currency composition for virtual currencies, the USA and the US dollar used to dominate the BitCoin market in the first years after its introduction. Nowadays, however, almost all BitCoin trading is done in China (Figure 4). The share of BitCoin that is traded via the China's mainland currency (yuan) jumped over the past few years, overtaking the US dollar as the dominant currency. From less than a 10% share in January 2012, the yuan now makes up nearly 100% of all BitCoin trading. Figure 4 shows the staggering rise of China as the dominant trader of BitCoin. In terms of the currency composition used for the trading of altcoins, it is clearly dominated by BitCoin – on average more than two thirds (68.17%) of all altcoin purchases are executed in the BitCoin currency, followed by the US dollar (14.19%), the Chinese yuan (14.63%), Euro (0.97%) and other altcoins (0.98%) (Table 1). The strong dominance of BitCoin as a medium of exchange in altcoin sales transactions suggests that there might be important interdependencies between BitCoin and altcoin prices.

*Price developments of BitCoin and Altcoins*

Figure 3 shows the price development of BitCoin and two altcoin price indices for the period 2012-2016. A visual inspection of BitCoin and altcoin price developments revel several commonalities between them. According to Figure 3, the prices of all virtual currencies have shown a significant fluctuation over time. The BitCoin price has increased from less than $20 value in 2012 to around $1100 at the end of 2013 while in the subsequent period it fluctuated between $200 and $800. In March 2017 the price of one BitCoin surpassed the spot price of an ounce of gold for the first time. Compared to major standard currencies, such as the US dollar or Euro, also altcoins show rather high price fluctuations (ALT19 and ALT100 in Figure 3).[3] First, altcoin prices decreased substantially over the period 2013-2016. The ALT19 index dropped from 100 on 13 December 2013 to 19 on 12 October 2016. The ALT100 index dropped from 1000 in January 2014 to around 90 in October 2016. This drop in altcoin prices appears to coincide with the BitCoin price decrease over the same period. Second, the prices of minor altcoins have decreased more substantially than the prices of major altcoins. This is visible from a stronger decline of the ALT100 index than the ALT19 index over the period 2013-2016. Note that the ALT100 index includes many minor altcoins, while ALT19 includes only major altcoins. Third, similarly to BitCoin, altcoins show a strong positive trend in the price development in the last year as depicted in Figure 3. Both ALT19 and ALT100 indices have increased by 55% and 124% in October 2016 compared to October 2015, respectively. This upward trend appears to be correlated with the BitCoin price increase over the same period (Figure 3). Hence, the price developments of BitCoin and altcoins seem to follow a similar pattern, which implies that the prices of both virtual currencies could be driven by the same external driver, or the prices of altcoins are following the price of the market leader (BitCoin).

*Hypothesis 2: Similarities in the BitCoin and Altcoin price formation mechanisms strengthen the market cointegration*

The second question that we aim to answer in this study is whether there are differences among individual altcoins in their price relation to BitCoin. Given that altcoin is not one homogenous currency, there exist important differences among altcoin and BitCoin characteristics. One can expect that differences in the virtual currencies' network and transaction setup and the supply/demand would affect the price formation asymmetrically and the extent to which altcoin prices follow the BitCoin market. Dissimilarities across virtual currencies may create differences in incentives (benefits and costs) for market participants (including network participants) leading to a differentiated price formation. In this section, we attempt to identify key virtual currency characteristics (supply, transaction execution/validation) that may be responsible for a stronger/weaker BitCoin and altcoin market cointegration. According to hypothesis 2, we expect that those alternative virtual currencies that in their price formation mechanisms are more similar to BitCoin would be stronger cointegrated and hence follow the BitCoin price development closer than other altcoins.

*Supply of virtual currencies*

An important determinant of any currency price is its supply in the short-, medium- and long-run. The higher is the growth rate of money in circulation, everything else equal, the higher would be inflationary pressures on the currency. Reversely, a weak supply of money growth may have deflationary impacts on the currency. As standard currencies, also virtual currencies regulate the total money supply by controlling both the stock of money in circulation and its growth rate. However, neither the stock nor the growth rate of money is controlled by a centralized financial authority or government, but by a software algorithm. They are exogenously pre-defined and publicly known to all virtual currency market participants from the time of the currency launch. This virtual currency feature contrasts standard currencies, where the supply of money is at the discretion of the Central Bank and thus not known a-priori (i.e. it depends on global macroeconomic developments and the monetary policy of the Central Bank) (Buchholz et al., 2012; Ciaian, Rajcaniova and Kancs, 2016a,b; Bouoiyour and Selmi, 2015).

---
[3] The ALT19 index is a composite price indicator of 19 altcoins, which includes 19 most important altcoins (such as Litecoin, Namecoin, NEM), whereas the ALT100 index is a composite price indicator of 100 different altcoins.

Many virtual currencies have followed BitCoin and have a fixed limit (maximum) of the total supply of coins that can be put in circulation (minted), e.g. Ripple, Litecoin and Dash (Table 2, Table 3). For these virtual currencies with a maximum supply, the growth rate of additionally minted coins is decreasing over time, it converges to zero when the maximum limit is reached. Different currencies with a fixed supply differ in terms of the speed, when the maximum limit is reached. Some virtual currencies reach the maximum stock faster than others. For example, Nxt, NEM and SuperNET have distributed all coins at the time when the currency was released, implying a zero-growth rate. Counterparty is an exception, it has a negative and decreasing growth of the coin supply, because a growing number of pre-defined number of coins are destroyed after each transaction.

Other virtual currencies have departed from the BitCoin supply model even further, by regulating their total stock of coins in circulation and/or the growth rate. For example, Ethereum, Monero, Dogecoin, Peercoin have opted for an unlimited total supply of coins (see Figure 5). Most virtual currencies with an unlimited supply impose an annual cap on newly minted coins. These currencies usually consider a fixed absolute number of coins that can be minted, meaning that the relative growth rate decreases over time.

Figure 5 and Figure 6 show the supply and development of virtual currency coins in circulation for currencies with a maximum and unlimited supply, respectively. There is a large heterogeneity in the development of the coin supply over time, with most of them having a continuous growth. Some others, for example, Ripple, MintCoin, BitShares and Nova coins exhibit a rather discontinuous development over time. In contrast, Nxt, NEM and SuperNET have a zero-growth rate (not shown in Figures), because all coins (the maximum supply) were distributed at the time when these currencies were launched.

Differences in the currency supply mechanism may result in differences in the currency price formation. Those virtual currencies that have followed BitCoin and have opted for a fixed supply are expected to be deflationary, if the currency demand increases over time. Further, a fixed coin supply may encourage a faster adoption of the virtual currency, as users and miners may have incentives to acquire coins immediately after their release in order to benefit from a potential future price increase.

Those virtual currencies that have departed from the BitCoin's supply mechanism and have opted for an unlimited supply of coins are expected to be inflationary particularly in the first years after their release, because the creation of new coins is large relative to the total stock of virtual currency coins already in circulation. In the long-run, however, they are expected to be deflationary, if their usage (demand) increases sufficiently high. That is, the cap on the absolute growth of new coins applied by these virtual currencies will represent a small share in relative terms, when the total stock in circulation becomes large, while a strong demand will push their prices up.

In summary, we would expect that, in terms of the underlying virtual currency supply mechanism, Ripple, Litecoin, Dash, Namecoin, Counterparty, Mintcoin, Qora and BitShares would exhibit a stronger relationship with the BitCoin price. Although, Nxt, NEM and SuperNET also have a fixed coin supply, the fact that they have released the total amount of coins right at the currency launch may have a differentiated impact on their price formation compared to BitCoin. The same applies to Counterparty which, in contrast to BitCoin, has a negative and decreasing growth of the total coin supply. Everything else equal, we would expect that the prices of Ethereum, Monero, Dogecoin, Peercoin, Novacoin, which have an unlimited supply of coins, would have a weaker relationship with the BitCoin price. Note that the supply is predefined and known by all market participants at the time of the virtual currency launch and at any point of time afterwards, implying that the supply is fully exogenous. Hence, we expect that the virtual currency supply would have only short-run effects on their prices.

*Demand for virtual currencies*

There can be identified two main types of demand for money/virtual currency: the transaction demand and speculative/asset demand.[4] The transaction demand for money/currency arises from the absence of a perfect synchronization of payments and receipts. Market participants may hold money/currency to bridge the gap between payments and receipts and to facilitate daily transactions. The transaction demand increases the money holding, which is expected to increase virtual currency prices. The speculative/asset demand is not driven by cash needs in transactions; instead, it stems from potential profit opportunities that may arise on financial markets and refers to cash held for the purpose of avoiding a capital loss from investments in financial assets such as bonds and stocks. A rise in the financial asset return (e.g. interest rate) causes their prices to fall, leading to a capital loss (negative return) from holding financial assets. Thus, investors may prefer to hold money/virtual currency to avoid losses from financial assets (Keynes, 1936; Nai Fovino et al., 2015). This implies a negative (positive) relationship between virtual currencies and financial asset returns/interest rate (financial asset prices) (e.g. Velde, 2013; Hanley, 2014; Yermack, 2014; Williams, 2014).

*Investment demand: competition among virtual currencies and other financial assets.* Potential investors consider virtual currencies as an alternative investment opportunity among many other possible investments (such as stocks, bonds, etc.). Given that virtual currencies are competing among each other for the attention of investors together with other financial assets and thus need to deliver a competitive return (compared to financial assets within the same risk class), the return arbitrage among alternative potential investment opportunities implies a positive price relationship between virtual currencies and financial assets. Real shocks to financial asset prices are expected to impact virtual currencies directly (both BitCoin and altcoins), if they are used as an investment asset rather than as money for transaction purposes (i.e. if the transaction demand is small). Hence, we would expect a direct (positive) relationship between virtual currencies and financial asset prices (Murphy, 2011). Thus, if the speculative/asset demand dominates compared to the transaction demand or, alternatively, if virtual currencies are perceived as an investment asset rather than as money for transactions, we would expect a direct relationship between virtual currencies and financial assets, whereas a negative relationship with the interest rate (Murphy, 2011).[5] For the same reasons, we may also expect a positive relationship between BitCoin and altcoins.

---

[4] In addition, there is also a precautionary demand for financial assets, such as money or virtual currency. It is the money that people hold in the case of emergency (Nai Fovino et al., 2015). However, given that the precautionary demand for virtual currencies is almost not-existent, it is not considered in the current analysis.

*Transaction demand: competition among different virtual currencies*. If instead the transaction demand dominates, the relationship between virtual currencies and financial assets is expected to be rather weak. In this case we may expect a negative relationship between virtual currencies and the exchange rate of standard currencies to be present. As noted by DeLeo and Stull (2014), if there is a transaction demand for virtual currencies, it implies individual's preferences for using virtual currencies as a medium of exchange for goods and services, as opposed to standard currencies. Since all currencies operate in free markets, where buyers and sellers come together to exchange a particular currency for goods and services, the transaction demand indicates how much demand there is for the particular currency compared to other currencies. Hence, competition between virtual currencies and standard currencies for transaction purposes (as a medium of exchange) implies a negative relationship. Preferred currencies used as medium of exchange for goods and services are expected to see their price rise at the expense of other (less preferred) currencies, whose price is expected to decrease. For the same reasons, we may also expect a negative relationship between BitCoin and altcoins.

*Information search costs.* Given that virtual currencies are relatively new currencies, their attractiveness for investment (asset) and/or transaction demand (and hence the aggregated demand) is affected by transaction costs of the information search for potential investors and users. According to Gervais, Kaniel, and Mingelgrin (2001); Grullon, Kanatas, and Weston (2004); Barber and Odean (2008) and Ciaian, Rajcaniova and Kancs (2016a), the information access (and thus attractiveness) to potential investors' decisions can be affected by an increase or decrease of attention in the mass media. The role of information is particularly important in the presence of many alternative investment opportunities (different virtual currencies and financial assets), positive search costs and security concerns. Given that the investment demand depends on the costs associated with searching for information for potential investment opportunities available on financial markets, such as, the stock exchange or virtual currencies, everything else equal, those investment opportunities which are under a particular attention in the mass media may be preferred by potential investors, because they reduce information search costs.[6]

The mass media attention-driven investment and use in transactions are expected to affect demand for virtual currencies either positively or negatively, depending on the type of news that dominate in the media at a given point of time. Note that all virtual currency prices are expected to be affected by the media attention. Given that BitCoin is the dominant virtual currency, its attention in the news media is expected to affect also altcoins' prices, besides their own media attention. Hence, the virtual media attention-driven demand for virtual currencies provides a further source of interdependencies between BitCoin and altcoins.

---

[5] Note that altcoin prices might be impacted by shocks to financial assets either directly or indirectly through BitCoin.

[6] Similar holds also for the information search about the cost of the payment method by virtual currency users. The choice of the payment method (e.g. PayPal, Visa, MasterCard, virtual currency) used for the exchange of goods and services depends on the costs associated with searching for information by potential users. Everything else equal, those payment methods that are under a particular attention in the mass media reduce search costs and hence may be preferred by users.

*Transaction execution/validation mechanism*

In order to execute virtual currency transactions, many altcoins have followed BitCoin and have adopted the Proof-of-Work (PoW) mechanism to secure and validate transactions (blocks) in the blockchain,[7] for example, Ethereum, Litecoin, Monero, Dash, Dogecoin and Namecoin (Table 2).[8] Miners that successfully complete the PoW receive a reward (new coins and transaction fees). The main principle of the PoW is that it is costly in terms of the computing power to produce but easy to verify by network participants. Miners must complete the PoW before their proposed block of transactions can be accepted by the network. That is, the PoW needs to be performed by a miner in order a new set of transactions (block) can be added to the distributed transaction database (block chain). The PoW is a random process; network computers need to solve a complex mathematical computation such that it satisfies certain pre-defined conditions. The difficulty of the computation (PoW) is continuously adjusted when new miners join the network or when miners invest in the computing power to ensure the established time interval (e.g. ten minutes for BitCoin) at which new blocks are generated. The success in completing the PoW is random, which makes it unpredictable which miner generates which block and thus receives the reward. Hence, the PoW has a built-in incentive mechanism that motivates network participants to invest in the computing power for mining in return of receiving transaction fees and newly minted coins while securely maintaining the whole system (Krawisz, 2013; Ali et al., 2014; Farell, 2015). A major disadvantage of the PoW mechanism is that it leads to a large investment in the computer power and the energy consumption with the only benefit to verify transactions (Farell, 2015).

Some other altcoins have opted for the Proof-of-Stake (PoS) mechanism for the block validation, which has a different reward system than the one used in BitCoin. The PoS is used, for example, by NxT and Qora. Under the PoS, priority is given to miners who hold a greater stake in the virtual currency. Miners that possess a larger amount of the virtual currency are prioritized compared to miners that possess less coins of the virtual currency. Hence, the probability that a miner succeeds in verifying a transaction and receiving the reward (new coins and/or transaction fees), is a function of the amount of virtual currency coins owned (and not of the computational power as in the case of the PoW).[9] The key challenge of the PoS system is the initial distribution of coins. While the PoW awards coins to miners who invest in the network (computing power), under the PoS, the coin distribution is rather ad-hoc and thus may lead to a fraudulent behavior among coin creators (miners).

---

[7] See Appendix 1 for a brief explanation of the blockchain technology.

[8] Namecoin is a spinoff of Bitcoin. Miners can mine Namecoin and Bitcoin coins at the same time without extra work while receiving rewards from both currencies.

[9] A modified version of PoS is the delegated PoS (DPoS) used by BitShares. Under DPoS in BitShares, stakeholders elect a number of witnesses who verify and add blocks to the blockchain. When witnesses produce a block, they are rewarded for their services. However, if witnesses fail to produce a block, they are not rewarded, and may be voted out by stakeholders. The advantage of DPoS is that it speeds up the confirmation procedure and thus increases the speed of transaction execution as well as it allows for a greater number of transactions to be included in a block compared to either PoW or PoS systems (Bitshares, 2017a,b).

A few currencies use a hybrid system of the PoS and PoW mechanisms, for example, Peercoin, Novacoin, Mintcoin (see Table 2). The hybrid system based on the PoS and PoW can address the problem of the initial coin distribution as the hybrid PoW/PoS system uses the PoW mechanism for the initial coin distribution to miners. Afterwards, the PoS mechanism gradually takes over the PoW mechanism (e.g. Peercoin, Mintcoin). Thus, the combination of the PoS and PoW avoids the initial distribution problem inherent to the PoS, while reducing the overall cost of the system characteristics for the PoW mechanism (Krawisz, 2013; Farell, 2015). For example, Mintcoin used the PoW during the first five weeks after its introduction, while afterwards it has almost completely switched to the PoS system.

Another mechanism used for the virtual currency transaction validation is the Proof-of-Importance (PoI) protocol used, for example, in NEM (see Table 2). The PoI aims to address the weakness of the PoS under which priority is given to miners with a high stake in the currency, which encourages coins' hoarding instead of using them in transactions. In the PoS protocol miners with small stakes have a lower chance of success for verifying transactions and receiving reward. Under the PoI system, miners' prioritization is based on the combination of two criteria: (i) stake in the virtual currency, and (ii) the intensity of miners' interactions and transactions with other users in the network (Beikverdi, 2015). Hence, the PoI encourages the use of virtual currencies but discourages hoarding by distributing rewards (transaction fees) based on the importance of each account in the network.[10] Thus a PoI-based virtual currency is expected to be more liquid when compared to a PoS-based currency, while compared to a PoW-based currency, it is expected to be less volatile.

The Byzantine Consensus Protocol (BCP) is a transaction verification mechanism that is based on the Byzantine Generals Problem initially developed by Lamport, Shostak and Pease (1982). In the Byzantine Generals experiment "… *several divisions of the Byzantine army are camped outside an enemy city, each division commanded by its own general. The generals can communicate with one another only by messenger. After observing the enemy, they must decide upon a common plan of action. However, some of the generals may be traitors, trying to prevent the loyal generals from reaching agreement. The generals must have an algorithm to guarantee that (a) all loyal generals decide upon the same plan of action ... and (b) a small number of traitors cannot cause the loyal generals to adopt a bad plan*." Distributed networks created by virtual currencies face an analogous problem, where participants need to decide whether other network participants are sending accurate messages (transactions). The BCP is used, for example, by Ripple (see Table 2). Various algorithms have been developed that provide solutions to the above problem. The consensus algorithm used in the BCP requires that 80 percent of network participants agree on validating each transaction. The advantage of this system is that the consensus can be fast and thus transactions can take place in seconds and is more energy efficient compared to several minutes required by the PoW systems as well as the mechanism is decoupled from the coin ownership as in the case of the PoS, thus avoiding the need for holding a stake in the virtual currency (Schwartz, Youngs, and Britto, 2014; Farell, 2015; Seibold and Samman, 2016).

---

[10] Majority of rewards (around 75%) in NEM are distributed based on PoI scores. The remaining rewards (around 25%) is allocated to marketing and development of the currency (Pangburn, 2014).

Yet another transaction verification mechanism used by some altcoins is the proof-of-burn (PoB) method. It is used, for example, by Counterparty (see Table 2). This method requires miners showing proof that they burned some digital coins during transaction (e.g. BitCoin) – they need to send them to a verifiably unspendable address – in return for receiving Counterparty coins.[11] The main advantage of the PoB is that it creates an equal opportunity for all users to obtain virtual currency coins.

SuperNET departs from other virtual currencies in that it is a basket of virtual currencies, similar to a closed-ended mutual fund. SuperNET aims to acquire 10 percent of the market capitalization of promising virtual currencies, i.e. innovative virtual currencies. Participants who donate to SuperNET receive tokens (coins) which are backed by virtual currencies comprising SuperNET (BitScan 2014).[12]

In the context of the present study, it is important to note that differences among transaction verification mechanisms adopted by different virtual currencies may be responsible for differences in the virtual currency price formation. For example, the PoW and PoS have different implications among others on the currency flexibility, market liquidity and price volatility. A PoS-based virtual currency (or DPoS) incentivizes the investment in the digital coin itself, whereas a PoW-based currency motivates investments in the underlying network (Krawisz 2013). According to Krawisz (2013), everything else equal, a PoW-based virtual currency will have a larger network with a higher capacity and a higher liquidity than the PoS network, whereas a PoS-based currency will have a greater price stability with lower incentives to contribute to blockchain.

The price behavior of PoI-based currencies is expected to be in between PoW- and PoS-based virtual currencies. Similar to PoS-based virtual currencies, a PoI-based currency is expected to have a greater price stability, because it incentivizes investment in the digital coin itself. On the other hand, PoI implies a greater liquidity, eventually leading to higher price responsiveness to market shocks (including BitCoin). Similarly, the price behavior of a BCP-based currency is expected to be in between PoW- and the PoS-based virtual currencies, because of a greater liquidity due to a relatively fast transaction execution and because it is decoupled from the coin ownership (there are no hoarding incentives). The price behavior of PoB-based currencies is expected to depend on the virtual currency that is required to be burned during the transaction verification process. For example, Counterparty requires burning BitCoins, hence its price is expected to be related stronger to the BitCoin price. In contrast, given that SuperNET is a basket of virtual currencies, its price is expected to be driven by the overall price development of virtual currencies, not necessarily only of the BitCoin price.

In summary, we would expect that the prices of Ethereum, Litecoin, Dash, Dogecoin, Monero, Namecoin and Counterparty, which as BitCoin are based on the PoW (or require burning BitCoins as in the case of Counterparty), would exhibit a stronger relationship with the BitCoin price than the prices of other altcoins. Given that other virtual currencies have adopted transaction verification mechanisms that are more distant from the PoW implemented in BitCoin, everything else equal, for these currencies we would expect a weaker relationship with the BitCoin price. For virtual currencies based on the PoS, we expect the smallest relationship with the BitCoin price (and other market shocks such as macro-financial variables), given that under this transaction verification mechanism their price is expected to show a greater price stability.

---

[11] The PoW lasted from 2$^{nd}$ January to 3$^{rd}$ February 2014.

[12] Additionally to the price of the UNITY coin, the SuperNET participants may be rewarded by payments similar to dividends generated from services (e.g. marketing) provided by SuperNET to the virtual currency community (BitScan, 2014).

## 3 Data sources and variable construction

*BitCoin and Altcoin prices*

In empirical estimations, we use daily data for the period 2013-2016 of BitCoin, 6 major altcoins and 10 minor altcoins. The selection of altcoins was based on the combination of three criteria: their differences and similarities in the currency implementation mechanisms compared to BitCoin and their virtual currency market share, as well as the data availability. The selected major altcoins are those that are listed in the top 10 virtual currencies in terms of their market capitalization (based on coinmarketcap.com). The selected minor altcoins have a comparably small market capitalization; they are not among top 10 virtual currencies. Further, we also consider two price indices - ALT19 and ALT100 - to capture the overall price development of altcoins and to provide a robustness check of the estimates for individual altcoins. The ALT19 index is a composite indicator of 19 altcoins, which includes 19 most important altcoins (such as Litecoin, Namecoin, NEM). It is calculated as a simple arithmetic sum of its 19 individual components. The ALT100 index comprises 100 different altcoins with the base value of 1000 set for 6 January 2014. Similar to ALT19, ALT100 is calculated as a simple arithmetic sum of 100 individual altcoins included in the index. Virtual currency prices are extracted from alt19.com, quandl.com, bitcoincharts.com and etherchain.org. The supply of virtual currencies is extracted from quandl.com and coinmarketcap.com (Table 4).

*Supply and demand of virtual currencies*

As explained in section 2, the supply of virtual currencies is exogenously pre-defined and publicly known to all virtual currency market participants from the time of the currency launch. For the purpose of this study we have extracted virtual currency supply and demand data from quandl.com and coinmarketcap.com (Table 4).

In order to account for the virtual media attention-driven demand, we include currency-specific Wikipedia page views in the estimation equation. According to Kristoufek (2013), the frequency of searches related to a virtual currency is a good proxy of the potential investors'/users' interest (search) in the particular currency. Wikipedia views are extracted from stats.grok.se and tools.wmflabs.org/pageviews. Note that there is no historical data available for the total demand of the studied virtual currencies (Table 4).

*Macroeconomic and financial developments*

To account for global macroeconomic and financial developments, we follow van Wijk (2013) and use the oil price, gold price, NASDAQ Composite[13] and 10-Year Treasury Constant Maturity Rate. The gold price is extracted from the World Gold Council (quandl.com), whereas oil price, NASDAQ Composite and the Treasury Rate data are extracted from the Federal Reserve Bank of St. Louis (Table 4).

---

[13] The NASDAQ Composite is a stock market index of the common stocks and similar securities listed on the NASDAQ stock market. It is among one of the three most-followed indices in USA stock markets.

In order to account for differences in the geographic/currency composition used for the altcoin trading, we include two exchange rates (USD/EUR and CNY/USD) in the estimation equation. According to Figure 4 and Table 1, BTC, USD, CNY and EUR are the most widely used currencies for the purchase of altcoins. The EUR/USD exchange rate data are extracted from the European Central Bank, while the USD/CNY exchange rate from the Federal Reserve Bank of St. Louis (Table 4).

**4 Estimation methodology**

The modelling approach used in this analysis is the Autoregressive Distributed Lag (ARDL) model proposed by Shin and Pesaran (1999), where first the cointegration space (i.e. long-run relationship) is estimated, after which we proceed by testing specific economic hypothesis within this space. Cointegration is a powerful tool for studying both the short- and long-run dynamics of BitCoin and altcoin supply and demand variables, and macroeconomic developments. This is because as long as there exists a cointegrating relationship among these variables, it means that there is a stationary long-run equilibrium relationship between individual non-stationary variables and in cases where these diverge from this long-run equilibrium, at least one of the variables in the system returns to the long-run equilibrium level (Juselius 2006).

In order to analyze the dynamic relationship between BitCoin and altcoin prices, we apply the ARDL bounds testing approach to cointegration developed by Pesaran and Shin (1999) and Pesaran et al. (2001). In the context of the present study, an important advantage of this approach is that it enables to estimate long- and short-run parameters simultaneously, it avoids endogeneity problems and can be applied irrespectively of whether underlying regressors are I(0), I(1) or mutually cointegrated (Pesaran and Shin, 1999). However, according to Ouattara (2004), if I(2) variables are present in the model the F statistics computed by Pesaran et al. (2001) become invalid, because the bounds test is based on the assumption that series should be either I(0) or I(1). Therefore, before applying the ARDL bounds test, we determine the order of integration of all variables using unit root tests, to make sure that none of the variable is integrated of order I(2) or beyond. In order to determine the stationarity of time series, we use three unit root tests: the augmented Dickey-Fuller (ADF) test, the Dickey-Fuller GLS test (DF-GLS) and the Zivot-Andrews test (ZA).

We apply both ADF and DF-GLS tests, as the DF-GLS test is considered to be a more efficient univariate test for autoregressive unit root recommended by Elliot, Rothenberg and Stock (1996). The test is a simple modification of the conventional augmented Dickey-Fuller (ADF) *t*-test, as it applies generalized least squares (GLS) detrending prior to running the ADF test regression. Compared with the ADF test, the DF-GLS test has the best overall performance in terms of the sample size and power. It "has substantially improved power when an unknown mean or trend is present" (Elliot, Rothenberg and Stock, 1996). However, as stated by Perron (1989), the existence of an exogenous shock which has a permanent effect will lead to a non-rejection of the unit root hypothesis even though the unit root is present. In order to take into account potential structural breaks in the series, which can lead to biased results when traditional tests are performed, in addition, we apply also the Zivot-Andrews test (Zivot and Andrews, 1992), allowing for one endogenously determined structural break. The Zivot and Andrews test takes into consideration structural breaks in intercept, trend or both. Testing for the unit root hypothesis allowing for a structural break can prevent the tests results to be biased towards a unit root and it can also identify when the structural break has occurred.

To investigate the presence of long-run relationships among series, we use the bound testing procedure of Pesaran, et al. (2001). The ARDL bound test is based on the Wald-test (F-statistic). The asymptotic distribution of the Wald-test is non-standard under the null hypothesis of no cointegration among variables. Two critical values are proposed by Pesaran et al. (2001) for the cointegration test. The lower critical bound assumes that all variables are I(0) meaning that there is no cointegration relationship between the examined variables. The upper bound assumes that all the variables are I(1) meaning that there is cointegration among the tested variables. When the computed F-statistics is greater than the upper bound critical value, then the null hypothesis of no long-run relationship is rejected. If the F-statistic is below the lower bound critical value, then the null hypothesis cannot be rejected (there is no cointegration among the examined variables). When the computed F statistics falls between the lower and upper bound, then the results are inconclusive.

The empirical representation of ARDL(*p,q,...,q*) model is the following:

(1) $\quad y_t = c_0 + c_1 t + \sum_{i=1}^{p} \emptyset_i y_{t-i} + \sum_{i=0}^{q} \beta_i x_{t-i} + \delta w_t + u_t$

where *p* is the number of lagged dependent variables, *q* specifies the number of lags for regressors, *t = max(p, q), ..., T*, for the sake of simplicity, we assume that the lag order *q* is the same for all variables in the *K × 1* vector $x_t$. $y_t$ is the dependent variable (virtual currency prices), $x_t$ denotes the independent variables, $w_t$ represents the exogenous variables, $u_t$ is a random "disturbance" term, $c_0$ and $c_1 t$ represent deterministic variables, intercept term and time trend, respectively.

In the second step, the long run relationship is estimated using the selected ARDL model. When there is a long run relationship between variables, there should exist an error correction representation. Therefore, the estimated error correction model is as follows:

(2) $\quad \Delta y_t = c_0 + c_1 t - \alpha(y_{t-i} - \theta x_{t-i}) + \sum_{i=1}^{p-1} \varphi_{yi} \Delta y_{t-i} + \omega \Delta x_t + \sum_{i=1}^{q-1} \varphi_{xi} \Delta x_{t-i} + \delta \Delta w_t + u_t$

where $\alpha$ is the speed of adjustment, $\theta$ are long-run coefficients, $\varphi$, $\omega$, $\delta$ are short-run coefficients to be estimated and $\Delta$ represents the differences.

A statistically significant coefficient, $\theta$, implies that there is a long-run relationship between variables (e.g. altcoin prices and BitCoin price and financial variables). The significant short run-coefficient coefficient shows that the corresponding variable has significant effect on the dependent variable (i.e. on virtual currencies prices) in the short-run. A series with significant both short-run and long-run coefficients indicates a strong causal effect on the dependent variable, while if only the short-run coefficient is significant, there is a weaker causal effect. Note that the supply of virtual currencies is exogenous in the estimated model (2) implying that there can be only a short-run causal effect of the virtual currencies' supply on their prices. The exogeneity assumption is motivated by the fact the supply is a-priori predefined and is known by all market participants and thus is independent of other variables (e.g. market conditions).

Moreover, because some prices might be exposed to external shocks, such as socio-political crises, currency devaluation, financial or economic crises, etc., dummy variables are included to account for their effect on the price development. According to Pesaran et al. (2001), the asymptotic theory of the ARDL approach is not affected by the inclusion of such dummy variables.

(3) $\Delta y_t = c_0 + \sum_{i=1}^{k} c_{D_i} DU_i + c_1 t - \alpha(y_{t-i} - \theta x_{t-i}) + \sum_{i=1}^{p-1} \varphi_{yi} \Delta y_{t-i} + \omega \Delta x_t + \sum_{i=1}^{q-1} \varphi_{xi} \Delta x_{t-i} + \delta \Delta w_t + u_t$

Dummy variables are defined by DU = 1 over the period t > τ$_i$ (τ$_i$ is the date that the shock occurred) and 0 elsewhere. We apply the Zivot Andrews procedure to price series in order to check the existence of breaks and the corresponding dates (Zivot and Andrews, 1992). Since each currency has its own market dynamics, break dates vary with currencies.

## 5 Results

Together with BitCoin, we study the price formation of 17 individual virtual currencies (BitCoin + 16 Altcoins) and two Altcoin price indices (ALT19 and ALT100), and by testing two hypotheses. Hypothesis 1 says that the prices of Altcoins are driven by the BitCoin price development. Hypothesis 2 says that similarities in the BitCoin and Altcoin price formation mechanisms strengthen the market cointegration. Hypothesis 1 suggests that all altcoins are expected to be affected by BitCoin, whereas Hypothesis 2 reinforces Hypothesis 1 for those altcoins that are more similar to BitCoin. We expect that BitCoin has a stronger and more significant impact on altcoins that are more similar to BitCoin compared to less similar altcoins. However, we may not be able to identify the two hypotheses in our estimations, because the effects of the two hypotheses might be cofounded (i.e. if the estimated BitCoin impacts are similar across all altcoins).

Table 5 summarizes the estimated models. We have estimated 4 models for each virtual currency, including for Altcoin price indices. In the estimated Altcoin models (M2.1-2.4), we alter the BitCoin price and ALT19 index (M2.1-2.2 *versus* M2.3-2.4) to test for the possible impact of both BitCoin and the general altcoin price level in determining individual altcoin prices. Similarly, we vary the coin supply of virtual currencies as well as the virtual media attention-driven demand for BitCoin and Altcoins in different models (M2.1-2.2 *versus* M2.3-2.4). We estimate similar specifications also for the BitCoin price. Following the previous literature on virtual currencies, we also control for global macro-financial developments in determining the prices of virtual currencies. In this section, we present estimation results, first the long-run, then the short-run.

### 5.1 Long-run relationship

Cointegration test results reveal that only half of virtual currencies are cointegrated, i.e. there exists a long-run equilibrium relationship between two or more variables: BitCoin, Ethereum, Litecoin, Dash, Peercoin, Namecoin, Nxt, Counterparty, Supernet and the aggregated altcoin price index (ALT100).

*Interdependencies between BitCoin and Altcoin prices*

ARDL test results confirm that there exists a long-run relationship between the BitCoin price and altcoin prices but only for few alternative virtual currencies (Table 6). Hence, contrary to expectations, our estimates do not support Hypothesis 1 that BitCoin would determine altcoin prices in the long-run. Further, our estimates do not fully support Hypothesis 2 that says that those altcoins that are more similar to BitCoin in their price formation mechanism (supply, transaction validation) follow the BitCoin price dynamics more closely than other alternative virtual currencies. Only the prices of Ethereum, Namecoin, NxT and SuperNET are found to be affected by the BitCoin price. Note that only Ethereum and Namecoin are based on the same PoW transaction validation mechanism as BitCoin. NxT applies the PoS mechanism, while SuperNET is a basket of alternative virtual currencies. In terms of the total coin supply, only Namecoin, NxT and SuperNET apply the maximum limit to the coin supply (as BitCoin does) whereas Ethereum has an unlimited coin supply (Table 3). For the rest of 12 individual altcoins, we do not find a long-run relationship between the BiCoin price and altcoin prices (Table 6). Hence, the BitCoin price is not included in the cointegrating space of these 12 altcoins. Five of them (i.e. Litecoin, Monero, Dash, Dogecoin, Novacoin) rely on the same PoW transaction validation mechanism as BitCoin, while the rest (i.e. Ripple, NEM, Peercoin, Counterparty, Mintcoin, Qora, BitShares) have a different validation mechanism than BitCoin. As for the currency supply mechanism, eight altcoins have a fixed supply as BitCoin (i.e. Ripple, Litecoin, Dash, NEM, Counterparty, Mintcoin, Qora, BitShares), whereas the rest have an unlimited coin supply (i.e. Monero, Dogecoin, Peercoin, Novacoin). These results imply that no long-run relationship exists between BitCoin and several altcoins that are similar to BitCoin on the one hand, as well as between BitCoin and several altcoins that are different from BitCoin on the other hand. Hence, generally we have to reject Hypothesis 2 that those altcoins that are more similar to BitCoin in their price formation mechanism follow the BitCoin price dynamics more closely.

We also find that altcoin price indices, ALT19 and ALT100, are not statistically significantly affected by real shocks to the BitCoin price in the long-run, which serves as a robustness check for individual currency estimates (Table 6). ALT19 and ALT100 indices represent a general development of altcoin prices, which does not seem to follow the BitCoin price. Although, BitCoin has a dominant position in the virtual currency market, general altcoin price developments do not appear to be linked to the BitCoin market. Similarly, the Altcoin price index itself affects only a small number of individual alternative virtual currencies: Ethereum, Peercoin, NxT and SuperNET (Table 6). The relationship with the BitCoin price is positive and statistically significant.

Whereas real shocks to the BitCoin price have a positive impact on a number of Altcoin prices (i.e. Ethereum, Namecoin, NxT and SuperNET), the relationship is inverse between the Altcoin price index and the BitCoin price. This is in contrast to the above presented results for individual altcoins (that are included in the cointegrating space), where the causal relationship between the altcoin price index and individual altcoin prices was positive (Table 6). This result can be explained by the competition effect between altcoins and BitCoin for the transaction and investment demand (see section 2), where users' adjustments in the altcoin transaction demand may cause an inverse response between the BitCoin price and the Altcoin price index. An increase in the demand for altcoins exercises an upward pressure on altcoin prices. In the same time, it takes away virtual currency market shares of BitCoin, hence exercising a downward pressure on the BitCoin price. Indeed, our estimation results show an inverse relationship between Altcoins and BitCoin; it is negative and statistically significant for both models of the ALT100 Altcoin price index, which suggest that positive shocks to the general altcoin price level – as captured by the Altcoin price index – affect the BitCoin price level negatively. These results are consistent and robust across all four estimated BitCoin models (not reported).

In summary, our estimation results suggest that in the long-run the BitCoin price has a direct impact only on prices of few Altcoins, implying a weak support for Hypothesis 1. In contrast, shocks to the Altcoin price index are transmitted inversely to the BitCoin price, whereas the BitCoin price does not have a statistically significant impact on Altcoin price indices. In line with Hypothesis 2, the prices of Ethereum, Namecoin, Nxt and Supernet, which are more similar to BitCoin in their price formation mechanisms (supply, demand, transaction validation), have a stronger cointegrating relationship with BitCoin than other virtual currencies. However, the price formation for other altcoins (for both with a similar and dissimilar formation mechanism as BitCoin) is not affected by BitCoin. Hence, contrary to Hypothesis 2, differences/similarities in virtual currency implementation mechanisms and supply specifics cannot fully explain the long-run relationship between BitCoin and altcoin prices.

*Virtual currency supply and demand*

As for the virtual currency supply, our results suggest that supply side variables are not cointegrated with virtual currency prices in the long-run. These results are in line with our expectations, as the virtual currency supply is pre-determined exogenously and publicly known to all market participants.

The variables capturing the impact of the virtual media attention-driven demand – views on Wikipedia – have a statistically significant impact on 6 virtual currency prices (BitCoin, Ethereum, Litecoin, Peercoin, Namecoin, NxT). Note that BitCoin views on Wikipedia are not impacting prices of most altcoins, which is an additional indirect evidence going against the Hypothesis 1. The prices of the rest of alternative virtual currencies are not affected by shocks in the virtual media attention-driven demand. As for the two Altcoin price indices, our estimates suggest that only ALT100 index is impacted by BitCoin and Altcoin views on Wikipedia.

The estimated weak impact of the media attention-driven demand on Altcoins could be explained by the fact that the type of information (rather basic) provided by Wikipedia becomes known for investors/users in the long-run. As a result, the number of Wikipedia queries by current/past investors about BitCoin and altcoins tends to decline over time (the variable views on Wikipedia becomes stationary), and it exercises a rather small impact on virtual currency prices in the long-run.

*Macroeconomic and financial developments*

Following the previous literature on virtual currencies, we also control for global macro-financial developments in determining the prices of virtual currencies. We find a somewhat stronger altcoin price interdependency with macro-financial developments, i.e. we find statistically more significant effects for a larger number of altcoins, compared to the BitCoin price. Our estimation results show that in the long-run at least two macro-financial variables have a statistically significant impact on 7 Altcoins (i.e. Litecoin, Dash, Peercoin, Namecoin, NxT, Counterparty and SuperNET) as well as on BitCoin. For remaining 9 altcoins (i.e. Ethereum, Ripple, Monero, NEM, Dogecoin, Novacoin, Mintcoin, Qora, BitShares) we do not find a statistically significant long-run relationship with any of the tested macro-financial variables. As for Altcoin price indices, our estimates suggest that the Alt100 index is affected by the gold price and the 10-Year Treasury Yield (Table 6). These results tend to suggest that macro-financial developments determine the altcoin price formation to a greater degree than BitCoin does in the long-run.

For virtual currencies with a statistically significant long-run relationship, financial asset-related variables such as the gold price and the 10-Year Treasury Yield have a mixed impact on their prices, while positive shocks to the NASDAQ index affect altcoin prices positively. The gold price has positive impact on prices of BitCoin, NxT and SuperNET, whereas negative on Dash and Namecoin. Similarly, the 10-Year Treasury Yield has a positive impact on BitCoin, Litecoin and Namecoin, whereas a negative impact on Dash and Counterparty. The NASDAQ index impacts positively the prices of BitCoin, Counterparty and SuperNET. Importantly, the estimated long-run relationship between the gold price and virtual currency is positive for most virtual currencies and estimated models with statistically significant effect. Only for two altcoins the estimated long-run relationship is negative - one for Dash and one for Namecoin.

The estimated positive long-run relationship between financial asset price indicators (NASDAQ Composite, gold price) and prices of virtual currencies is in line with the transactions demand and/or investment/asset demand theory of money (see section 2). Higher financial asset prices stimulate the holding of virtual currencies (i.e. positive relationship) to avoid the risk of losses from a potential price decrease of holding other financial assets (e.g. stocks). If virtual currencies are perceived as investment assets, the positive long-run relationship may be a result of the arbitrage return between alternative investment opportunities (including virtual currencies). Similarly, a higher expected bonds' interest rate may induce market participants to hold virtual currencies to avoid the loss from a fall in the bond price, implying an inverse relation (negative) between the 10-Year Treasury Yield and virtual currencies prices.

We do not find any cointegrating relationships between the long-run crude oil price and most virtual currency prices. According to our estimates, real shocks to the oil price are transmitted only to the long-run prices of Peercoin (only in one estimated model). These results are consistent with the findings of Ciaian, Rajcaniova and Kancs (2016a). Thus, the oil price does not seem to determine virtual currency prices in the long-run.

The importance of exchange rates in driving virtual currencies' prices is similar to the estimates reported above for financial asset indices. According to our long-run estimates, exchange rates have a significant long-run impact on 7 virtual currency markets (i.e. BitCoin, Litecoin, Peercoin, Namecoin, NxT, Counterparty and SuperNET). However, the estimated impact of USD/EUR and CNY/USD exchange rates is differentiated across virtual currencies. The USD/EUR exchange rate affects 3 virtual currency markets (i.e. BitCoin, Namecoin, Counterparty), while the CNY/USD exchange affects 7 virtual currencies (Table 6). This could be explained by the higher uncertainty linked to Chinese financial markets, hence causing a greater dependency between CNY/USD and virtual currencies. For some virtual currencies, a positive USD/EUR shock is transmitted into a virtual currency appreciation, e.g. BitCoin, for other virtual currencies, into a currency depreciation, e.g. Supernet. In the case of a positive CNY/USD shock, it is transmitted into an appreciation of e.g. Peercoin, whereas into a depreciation of e.g. Counterparty. These findings fall within our expectations, as there are important differences in the currency composition that are used for the BitCoin and altcoin trading (Table 6). The negative relationship estimated for exchange rates and virtual currencies could be explained by the competition between currencies for their use in transactions (as a medium of exchange). However, for several currencies we find a positive relationship between exchange rates and virtual currency prices, suggesting that various other market-specific factors may confound our estimations, e.g. the inflation rate, resulting in differentiated interdependencies between exchange rates and virtual currencies. As for Altcoin price indices, cointegration test results reject a long-run relationship, implying that real shocks to USD/EUR and CNY/USD exchange rates are not transmitted to the general Altcoin price level.

In summary, our results suggest that macro-financial developments drive virtual currencies to a larger extent than the BitCoin price. However, their impact is not overwhelming and strongly differentiated across virtual currencies. Moreover, several virtual currencies and Altcoin price indices do not seem to be significantly impacted by global macro-financial developments.

## 5.2 Short-run dynamics

*Interdependencies between BitCoin and Altcoin prices*

Generally, short-run results are more significant across virtual currencies though also more heterogeneous than long-run results, both in terms of signs and dynamics (Table 7). As for Hypothesis 1, we find that in the short-run the BitCoin price consistently impacts most of altcoins (15). Dash is the only altcoin for which the impact of BitCoin is not statistically significant. These results suggest that the BitCoin dominance is reflected in short-run Altcoin price adjustments rather than is the long-run joint BitCoin-Altcoin price relationship. This is a weaker form of the price interdependency. It implies that although, some of exogenous shocks, e.g. macro-financial developments, are associated with long-run effects on the price variability, a significant source of price shocks in BitCoin and altcoin markets are associated with short-run disruptions. These results suggest that for most alternative virtual currencies we cannot reject Hypothesis 1 for the short-run. The BitCoin price has a positive and statistically significant impact on altcoin prices. The same applies also for the altcoin price index which is our robustness check for the results obtained for individual altcoins. In contrast, positive shocks to the altcoin price index have a negative impact on the BitCoin price, which is consistent with our long-run estimation results presented above.

According to Hypothesis 2, those altcoins that are more similar in their price formation mechanism to BitCoin would follow the BitCoin price development more closely. According to the results reported in Table 7, apart from Dash, all other virtual currencies that follow the same/similar price formation mechanisms as BitCoin (Table 3) are impacted by shocks to the BitCoin market for several periods: Litecoin 3 and Namecoin 3. This is significantly higher than the average number of impacted lags for altcoins, which is 1.9 (first row in Table 7). In the same time, we note that there are also other altcoins with a different price formation mechanism that are significantly impacted by the shocks to the BitCoin price, e.g. Ripple (4), Novacoin (3), BitShares (3).

Overall, these estimates do not allow us to fully identify the effects of the two Hypotheses separately. Based on these results, we cannot fully reject Hypothesis 2. However, given that BitCoin consistently impacts all altcoins, these results suggest that Hypothesis 1 tends to dominate over Hypothesis 2. Further, there might be also other confounding factors, e.g. global macro-financial developments that might be transmitted through BitCoin into altcoin prices witch corroborates with Hypothesis 1.

*Virtual currency supply and demand*

The short-run impact of media attention-driven demand on virtual currencies' prices is considerably stronger than in the long-run. Most virtual currencies are impacted by Wikipedia views in the short-run, either by their own Wikipedia views or the combined altcoin Wikipedia views (the exception is Qora) (Table 7). Currency-own Wikipedia views have a positive and statistically significant impact on 9 altcoins in the short-run. Similarly, combined altcoin Wikipedia views impact 10 altcoins in the short-run. BitCoin Wikipedia views have a marginal impact on altcoin prices; they impact only four altcoins.

These estimates suggest that own attention-driven demand (information search) is more important in determining altcoin prices than the information search for other virtual currencies. That is, own information search is reflected in short-run Altcoin price adjustments. Further, our results show that BitCoin information search is marginally transmitted on altcoin short-run price changes.

As for supply-side variables, the currency supply (as captured by the amount of coins in circulation) does not have a statistically significant impact on most virtual currency prices. Only 4 virtual currencies are impacted by the supply. These results can be explained by the fact that the virtual currency supply is an exogenous variable, as it is pre-determined a priori and publicly known to all market participants. Hence, coin supply developments are likely incorporated in the expectations of market participants, they do not generate short-run price deviations from the long-run equilibrium situation for most virtual currencies.

*Macroeconomic and financial developments*

The impact of global macroeconomic and financial developments on altcoin prices is rather significant in the short-run. Generally, this confirms our long-run results. However more virtual currencies are impacted by macro-financial indicators in the short-run than in the long-run. According to the results in Table 7, the macro-financial indicators impact 13 virtual currencies as well as altcoin indices. Similar to long-run estimation results, the two exchange rates (USD/EUR and CNY/USD) have a significant impact on prices of most altcoins. However, whereas in the long-run the CNY/USD exchange rate is more important, in the short-run the USD/EUR exchange rate affects altcoin prices more significantly (11 altcoin prices/indices are impacted by USD/EUR in the short-run). The prices of most altcoins are impacted also by real shocks to the gold and oil prices. Note that the impact of the latter was not statistically significant in the long-run. The 10-Year Treasury Constant Maturity Rate and the NASDAQ Composite affect altcoin prices less in the short-run. Overall, the short-run impact of global macro-economic developments on altcoin and BitCoin prices seem to be stronger than the impact of supply and demand variables. These results are consistent with findings of van Wijk (2013).

# 6 Conclusions

Since around a decade, virtual currencies represent a new phenomenon on global financial markets. By providing an alternative money, they function outside centralized financial institutions. While offering a less expensive alternative to mainstream currencies in terms of transaction costs, the prices of virtual currencies are developing considerably more erratically and fluctuate much wider than those of standard currencies. Despite the comparably high market volatility, there is little known about their price formation mechanisms and altcoin interdependencies with the BitCoin market. Indeed, there are good reasons to believe that BitCoin and altcoin prices might be interdependent, given that BitCoin is the dominant virtual currency, similar patterns in BitCoin and altcoin price developments, as well as the fact that a large majority of altcoin purchases are executed in BitCoins.

The present paper attempts to fill this knowledge gap by studying price interdependencies between BitCoin and altcoin markets in the short- and long-run. In particular, we test two hypotheses related to BitCoin and altcoin markets and their interdependencies. Hypothesis 1 says that the prices of Altcoins are driven by the BitCoin market development. Hypothesis 2 says that similarities in the BitCoin and Altcoin price formation mechanisms strengthen the market cointegration. Following the previous literature on virtual currencies, we also control for global macro-financial developments in determining the prices of virtual currencies.

To answer these questions, we apply time-series analytical mechanisms to daily data of 17 virtual currencies (BitCoin + 16 Altcoins) and altcoin price indices for the period 2013-2016. Following the previous literature, we also include demand and supply variables, and development of global macro-financial indicators.

Our empirical findings for Hypothesis 1 suggest that BitCoin and altcoin prices are indeed interdependent, though more in the short- than in the long-run. In the short-run, 15 altcoin prices (out of 16 studied altcoins) are impacted by real shocks to BitCoin prices. In the long-run, however, only 4 altcoins are cointegrated with BitCoin prices. Thus, we cannot reject Hypothesis 1 for the short-run, while we reject Hypothesis 1 for the long-run BitCoin-Altcoin relationship.

The empirical results for Hypothesis 2 suggest that among altcoins with the strongest price interdependencies with BitCoin there are both similar and dissimilar virtual currencies with respect to the price formation of BitCoin. In the short-run, the number of periods where BitCoin price shocks are transmitted to altcoin prices is indeed higher for those altcoins that have the same/similar price formation mechanism as BitCoin. However, we also find a relatively strong short-run impact of BitCoin for some altcoins that are dissimilar to BitCoin. In the long-run, however, the differences/ similarities of altcoins with BitCoin cannot explain the long-run relationship between BitCoin and altcoin prices. Thus, we cannot fully identify Hypothesis 2 for the short-run, but we reject Hypothesis 2 for the long-run BitCoin-Altcoin relationship. However, given that BitCoin consistently impacts all altcoins in the short-run, this suggests that Hypothesis 1 tends to dominate over Hypothesis 2 for the short-run BitCoin-altcoin relationship.

As regards the virtual currency supply and demand, their impact on virtual currency prices (both BitCoin and altcoins) is stronger in the short- than in the long-run. In contrast, the impact of global macroeconomic and financial developments on virtual currency prices is statistically significant both in the short- and long-run for most virtual currencies. Particularly, the gold price, the two exchange rates (USD/EUR and CNY/USD) and the 10-Year Treasury Constant Maturity Rate have a statistically significant impact on BitCoin and altcoin prices in the short- and long-run. Further, our results suggest that in the long-run macro-financial indicators determine the altcoin price formation to a greater degree than BitCoin does. In the short-run, macro-financial indicators and BitCoin appear to impact altcoin prices equally.

Our findings have important messages both for virtual currency users and investors. First, the dominant virtual currency – BitCoin – has only a limited impact on the prices of altcoins in the long-run. This implies, for example, it is unlikely that the currently peaking BitCoin price will drive up the prices of altcoins in the long-run. Second, global macroeconomic and financial developments determine virtual currency prices to a larger extent than virtual currency-specific factors, such as the currency supply and demand. Hence, the price development of virtual currencies is intrinsically difficult to predict in the long-run. Third, although BitCoin plays a small role in the long-run price formation, it the short-run it may have important implications for altcoin prices. A significant source of the short-run price fluctuations of altcoins might be associated with BitCoin disruptions. Fourth, our findings suggest that the empirical evidence about the altcoin price formation and its relation to BitCoin is important not only for virtual currency users and potential investors, but it also provides an interesting laboratory for studying interdependencies in the currency price formation in the short- and long-run.

**Table 1. Currency composition of the global altcoin trading volume (%)**

|  | BTC | USD | CNY | EUR | Altcoin | Others |
|---|---|---|---|---|---|---|
| Ethereum (ETH) | 47.01 | 26.13 | 2.31 | 8.28 |  | 16.27 |
| Litecoin (LTC) | 40.74 | 22.59 | 33.69 | 2.57 |  | 0.41 |
| DogeCoin (DOGE) | 57.75 |  | 41.68 |  |  | 0.57 |
| Monero (XMR) | 84.40 | 11.44 |  | 4.16 |  | 0.00 |
| Ripple (XRP) | 84.27 | 7.11 | 5.20 | 3.42 |  | 0.00 |
| DigitalCash (DASH) | 78.95 | 18.10 | 1.26 |  |  | 1.69 |
| NEM (XEM) | 76.72 |  | 23.28 |  |  | 0.00 |
| PeerCoin (PPC) | 58.37 | 28.33 | 13.22 |  |  | 0.08 |
| Bitshares (BTS) | 71.87 |  | 28.13 |  |  | 0.00 |
| Nxt (NXT) | 49.78 | 3.74 | 46.42 |  |  | 0.06 |
| NameCoin (NMC) | 27.74 | 71.05 | 1.21 |  |  | 0.00 |
| NovaCoin (NVC) | 18.93 | 81.07 |  |  |  | 0.00 |
| CounterParty (XCP) | 99.77 |  |  |  |  | 0.23 |
| Qora (QORA) | 99.79 |  |  |  |  | 0.21 |
| MintCoin (MINT) | 84.23 |  |  |  | 15.77 | 0.00 |
| Bitshares (BTS) | 72.42 |  | 27.58 |  |  | 0.00 |
| FeatherCoin (FTC) | 93.84 |  | 5.70 |  |  | 0.46 |
| PrimeCoin (XLB) | 51.65 |  | 48.35 |  |  | 0.00 |
| Lisk (LSK) | 97.06 |  |  |  | 2.90 | 0.04 |
| **ALT19** | **68.17** | **14.19** | **14.63** | **0.97** | **0.98** | **1.05** |

Source: htttps://www.cryptocompare.com/coins/

**Table 2. Characteristics of selected virtual currencies**

| | Symbol | Date of release | Duration of a block creation | Blocks generation mechanism | Growth of supply | Maximum supply |
|---|---|---|---|---|---|---|
| *Major VC* | | | | | | |
| BitCoin | BTC | 2009 | 10 minutes | PoW | Decreasing rate (halved every 210,000 blocks) | 21 Million |
| Ethereum | ETH | 2015 | 15 to 17 seconds | PoW** | Smoothly decreasing (in relative terms) (when reaching 72 million units, the supply will stay at max 18 million of new coins per year) | Unlimited |
| Ripple | XRP | 2012 | 3-5 seconds | BC | Undefined: half of all units will be released for circulation, while OpenCoin* will retain the rest. | 100 billion |
| Litecoin | LTC | 2011 | 2.5 minutes | PoW | Decreasing rate (halved every 840,000 blocks) | 84 million |
| Monero | XMR | 2014 | 2 minutes | PoW | Smoothly decreasing (in relative terms) (when reaching 18.4 million units, the supply will stay constant at 0.6 new coins per 2-minutes block) | Unlimited |
| Dash | DASH | 2015 | 2.5 minutes | PoW | Decreasing rate (newly minted coins decrease 7.1% annually) | 22 million |
| NEM | | 2015 | 1 minute | PoI | Zero (100% of coins were distributed when launched) | 9 billion |
| *Minor VC* | | | | | | |
| Dogecoin | DOGE | 2013 | 1 minute | PoW | Smoothly decreasing rate (in relative terms) (when reaching 100 billion units, the supply will stay constant at 5.256 billion of new coins per year) | Unlimited |
| Peercoin | PPC | 2012 | 10 minutes | Hybrid (PoS & PoW) | Semi-constant (the supply increases at a rate of up to 1% per year but partially offset by destruction of 0.01 PPC per transaction) | Unlimited |
| Namecoin | NMC | 2011 | 10 minutes | PoW | Decreasing (halved every 4 years) | 21 Million |
| Novacoin | NVC | 2013 | 10 minutes | Hybrid (PoS & PoW) | dynamic inflation | Unlimited |
| NxT | NXT | 2013 | 1 minute | PoS | Zero (100% of coins were distributed when launched) | 1 billion |
| Counterparty | XCP | 2014 | 10 minutes | PoB | Negative growth at decreasing rate (coins are destroyed) | 2.6 million |
| Mintcoin | MINT | 2014 | 30 seconds | Hybrid (PoS & PoW) | Decreasing rate (PoS: first year: 20%; second year: 15%; third year: 10%; fourth year and after: 5%. PoW reward: halved every week in first 5 weeks, 1 coin per block afterwards) | 70 billions |
| Qora | QORA | 2014 | 1-5 minute | PoS | Undefined | 10 bullion |
| SuperNET | UNITY | 2014 | - | Basket of virtual currencies | Zero | 816 061 |
| BitShares | BTS | 2014 | 5-10 seconds | DPoS | Variable | 3.7 Billion |

Sources: Bitshares (2016); Brown 2013; CoinDesk (2014), Bovaird (2016); Johnson (2016)
Notes; Proof-of-Work: PoW; Proof-of-Stake: PoS; Delegated Proof-of-Stake: DPoS; Proof-of-Importance: PoI; Byzantine Consensus: BC; Proof-of-burn: PoB;* OpenCoin is the company that developed Ripple; ** Ethereum plans to move to PoS protocol.

**Table 3. Summary description of similarities of altcoins with Bitcoin**

|  | Coin supply | | Validation mechanism | Overall |
|---|---|---|---|---|
|  | Maximum supply | Growth of supply |  |  |
| Ethereum | No | No | Yes | No (1/3) |
| Ripple | Yes | Yes | No | No (2/3) |
| Litecoin | Yes | Yes | Yes | Yes |
| Monero | No | No | Yes | No (1/3) |
| Dash | Yes | Yes | Yes | Yes |
| NEM | Yes | No | No | No (1/3) |
| Dogecoin | No | No | Yes | No (1/3) |
| Peercoin | No | No | No | No (0/3) |
| Namecoin | Yes | Yes | Yes | Yes |
| Novacoin | No | No | No | No (0/3) |
| NxT | Yes | No | No | No (1/3) |
| Counterparty | Yes | Yes | No | No (2/3) |
| Mintcoin | Yes | Yes | No | No (2/3) |
| Qora | Yes | Yes | No | No (2/3) |
| SuperNET | Yes | No | No | No (1/3) |
| BitShares | Yes | Yes | No | No (2/3) |

Notes: Yes: Altcon is similar to BitCoin; No: Altcon is not similar to BitCoin

**Table 4. Data sources**

| Variable | Unit | Variable name | Frequency | Period | Source |
|---|---|---|---|---|---|
| *Prices of VC* | | | | | |
| bitcoin_usd | USD per 1 unit | BitCoin price | Daily | 13 Jul 2012 - 12 Oct 2016 | quandl.com |
| ethereum_usd | USD per 1 unit | Ethereum price | Daily | 30 Aug 2015 - 12 Oct 2016 | etherchain.org |
| ripple_usd | USD per 1 unit | Ripple price | Daily | 4 Jan 2014 - 19 Jul 2016 | bitcoincharts.com |
| litecoin_usd | USD per 1 unit | Litecoin price | Daily | 13 Jul 2012 - 12 Oct 2016 | quandl.com |
| monero_usd | USD per 1 unit | Monero price | Daily | 10 Mar 2015 - 12 Oct 2016 | quandl.com |
| dash_usd | USD per 1 unit | Dash price | Daily | 25 Mar 2015 - 12 Oct 2016 | quandl.com |
| nem_usd | USD per 1 unit | NEM price | Daily | 13 Apr 2015 - 12 Oct 2016 | quandl.com |
| dogecoin_usd | USD per 1 unit | Dogecoin price | Daily | 22 Mar 2014 - 12 Oct 2016 | quandl.com |
| peercoin_usd | USD per 1 unit | Peercoin price | Daily | 1 Apr 2014 - 12 Oct 2016 | alt19.com |
| namecoin_usd | USD per 1 unit | Namecoin price | Daily | 1 Apr 2014 - 12 Oct 2016 | alt19.com |
| novacoin_usd | USD per 1 unit | Novacoin price | Daily | 1 Apr 2014 - 12 Oct 2016 | alt19.com |
| nxt_usd | USD per 1 unit | NxT price | Daily | 14 Apr 2014 - 12 Oct 2016 | quandl.com |
| counterparty_usd | USD per 1 unit | Counterparty price | Daily | 14 Apr 2014 - 12 Oct 2016 | quandl.com |
| mintcoin_usd | USD per 1 unit | Mintcoin price | Daily | 14 Apr 2014 - 12 Oct 2016 | quandl.com |
| qora_usd | USD per 1 unit | Qora price | Daily | 27 Jun 2014 - 12 Oct 2016 | quandl.com |
| supernet_usd | USD per 1 unit | SuperNET price | Daily | 24 Sep 2014 - 12 Oct 2016 | quandl.com |
| bitshares_usd | USD per 1 unit | BitShares price | Daily | 13 Nov 2014 - 12 Oct 2016 | quandl.com |
| alt19 | Index in BTC | Price index of 19 altcoins | Daily | 13 Dec 2013 - 12 Oct 2016 | alt19.com |
| alt100usd | Index in USD | Price index of 100 altcoins | Daily | 7 Jan 2014 - 12 Oct 2016 | alt19.com |
| *Supply of VC* | | | | | |
| supply_bitcoin | No. of units | BitCoin supply | Daily | 13 Jul 2012 - 12 Oct 2016 | quandl.com |
| supply_ethereum | No. of units | Ethereum Supply | Weekly | 9 Aug 2015 - 12 Oct 2016 | coinmarketcap.com |
| supply_ripple | No. of units | Ripple Supply | Weekly | 11 Aug 2013 - 12 Oct 2016 | coinmarketcap.com |
| supply_litecoin | No. of units | Litecoin Supply | Weekly | 28 Apr 2013 - 12 Oct 2016 | coinmarketcap.com |
| supply_monero | No. of units | Monero Supply | Weekly | 25 May 2014 - 12 Oct 2016 | coinmarketcap.com |
| supply_dash | No. of units | Dash Supply | Weekly | 16 Feb 2014 - 12 Oct 2016 | coinmarketcap.com |
| supply_nem | No. of units | Nem Supply | Weekly | 6 Apr 2015 - 12 Oct 2016 | coinmarketcap.com |
| supply_dogecoin | No. of units | Dogecoin Supply | Weekly | 22 Dec 2013 - 12 Oct 2016 | coinmarketcap.com |
| supply_peercoin | No. of units | Peercoin Supply | Weekly | 28 Apr 2013 - 12 Oct 2016 | coinmarketcap.com |
| supply_namecoin | No. of units | Namecoin Supply | Weekly | 28 Apr 2013 - 12 Oct 2016 | coinmarketcap.com |
| supply_novacoin | No. of units | Novacoin Supply | Weekly | 28 Apr 2013 - 12 Oct 2016 | coinmarketcap.com |
| supply_nxt | No. of units | Nxt Supply | Weekly | 8 Dec 2013 - 12 Oct 2016 | coinmarketcap.com |
| supply_counterparty | No. of units | Counterparty Supply | Weekly | 16 Feb 2014 - 12 Oct 2016 | coinmarketcap.com |
| supply_mintcoin | No. of units | Mintcoin Supply | Weekly | 203 Feb 2014 - 12 Oct 2016 | coinmarketcap.com |
| supply_qora | No. of units | Qora Supply | Weekly | 1 Jun 2014 - 12 Oct 2016 | coinmarketcap.com |
| supply_supernet | No. of units | Supernet Supply | Weekly | 5 Oct 2014 - 12 Oct 2016 | coinmarketcap.com |
| supply_bitshares | No. of units | Bitshares Supply | Weekly | 27 Jul 2014 - 12 Oct 2016 | coinmarketcap.com |
| supply_altcoins | No. of units | Sum of Altcoin supply | Weekly | 23 Feb 2014 - 12 Oct 2016 | Calculated |
| *Virtual media attention-driven demand* | | | | | |
| wiki_bitcoin | No. of views | BitCoin | Daily | 13 Jul 2012 - 12 Oct 2016 | SG & TW |
| wiki_ethereum | No. of views | Ethereum | Daily | 13 Jul 2012 - 12 Oct 2016 | SG & TW |
| wiki_ripple | No. of views | Ripple (payment protocol) | Daily | 1 Dec 2013 - 12 Oct 2016 | SG & TW |
| wiki_litecoin | No. of views | Litecoin | Daily | 13 Jul 2012 - 12 Oct 2016 | SG & TW |
| wiki_monero | No. of views | Monero (cryptocurrency) | Daily | 1 Mar 2015 - 12 Oct 2016 | SG & TW |
| wiki_dash | No. of views | Dash (cryptocurrency) | Daily | 1 Apr 2015 - 12 Oct 2016 | SG & TW |
| wiki_nem | No. of views | NEM (cryptocurrency) | Daily | 1 May 2015 - 12 Oct 2016 | SG & TW |
| wiki_dogecoin | No. of views | Dogecoin | Daily | 1 Dec 2013 - 12 Oct 2016 | SG & TW |
| wiki_peercoin | No. of views | Peercoin | Daily | 1 Apr 2013 - 12 Oct 2016 | SG & TW |
| wiki_namecoin | No. of views | Namecoin | Daily | 13 Jul 2012 - 12 Oct 2016 | SG & TW |
| wiki_altcoins | No. of views | Sum of altcoins | Daily | 1 Dec 2013 - 12 Oct 2016 | Calculated |
| *Macro-financial variables* | | | | | |
| gold_price | USD per 1 ounce | Gold price | Daily | 13 Jul 2012 - 12 Oct 2016 | WGC, quandl.com |
| nasdaq | Index in USD | NASDAQ Composite Index | Daily | 13 Jul 2012 - 12 Oct 2016 | FRED |
| treasury_rate10y | % rate | 10-Year Treasury Constant Maturity Rate | Daily | 13 Jul 2012 - 12 Oct 2016 | FRED |
| e_usd_eur | USD per 1 EURO | EUR/USD exchange rate | Daily | 13 Jul 2012 - 12 Oct 2016 | ECB |
| e_yuan_usd | CNY per 1 USD | USD/CNY exchange rate | Daily | 13 Jul 2012 - 12 Oct 2016 | FRED |
| oil_price | Dollars per Barrel | Crude Oil Prices: Brent (DCOILBRENTEU) | Daily | 13 Jul 2012 - 12 Oct 2016 | FRED |

Notes: FRED: Federal Reserve Economic Data, Federal Reserve Bank of St. Louis; ECB: European Central Bank; SG & TW: stats.grok.se and tools.wmflabs.org/pageviews, WGC: World Gold Council.

**Table 5. Specification of the estimated models**

|  | BitCoin | | | | Altcoins | | | |
|---|---|---|---|---|---|---|---|---|
|  | M 1.1 | M 1.2 | M 1.3 | M 1.4 | M 2.1 | M 2.2 | M 2.3 | M 2.4 |
| *Prices* | | | | | | | | |
| bitcoin_usd | | | | | X | X | | |
| alt19 | | X | | X | | | X | X |
| alt100usd | X | | X | | | | | |
| *Supply of virtual currencies* | | | | | | | | |
| supply_bitcoin | X | X | X | | X | X | | |
| supply_own (for individual altcoins) | | | | | X | | | X |
| supply_altcoins_total | X | X | X | X | | | X | X |
| *Media attention-driven demand* | | | | | | | | |
| wiki_bitcoin | X | X | | | X | X | | |
| wiki_own (for individual altcoins) | | | | | X | | | X |
| wiki_altcoins_total | X | X | X | X | | | X | X |
| *Macro-financial variables* | | | | | | | | |
| gold_price | X | X | X | X | X | X | X | X |
| nasdaq | X | X | X | X | X | X | X | X |
| treasury_rate10y | X | X | X | X | X | X | X | X |
| e_usd_eur | X | X | X | X | X | X | X | X |
| e_yuan_usd | X | X | X | X | X | X | X | X |
| oil_price | X | X | X | X | X | X | X | X |

**Table 6. Estimation results: long-run relationships between virtual currency prices**

| | BTC | ETH | XRP | LTC | XMR | DASH | NEM | DOGE | PPC | NMC | NVC | NXT | XCP | MINT | QORA | UNITY | BTS | alt19 | alt100usd |
|---|---|---|---|---|---|---|---|---|---|---|---|---|---|---|---|---|---|---|---|
| *Prices* | | | | | | | | | | | | | | | | | | | |
| bitcoin_usd | | (+)** | | | | | | | | (+)** | | (+)*** | | | | (+)** | | | |
| alt19 | (-)** | (+)*** | | | | | | | (+)*** | | | (+)** | | | | (+)** | | | |
| alt100usd | (-)** | | | | | | | | | | | | | | | | | (+)** | |
| *Macro-financial variables* | | | | | | | | | | | | | | | | | | | |
| gold_price | (+)** | | | | (-)* | | | | | (-)* | | (+)** | | | | (+)** | | | (+)* |
| nasdaq | (+)* | | | | | | | | | | | | (+)*** | | | (+)** | | | |
| treasury_rate10y | (+)* | | | (+)*** | (-)** | | | | | (+)* | | | (-)* | | | | | | (+)* |
| e_usd_eur | (+)* | | | | | | | | | (+)** | | | (+)*** | | | | | | |
| e_yuan_usd | (+)** | | | (+)*** | | | | | (+)*** | (+)*** | | (+)*** | (-)** | | | (-)*** | | | |
| oil_price | | | | | | | | | (+)** | | | | | | | | | | |
| *Wikipedia views* | | | | | | | | | | | | | | | | | | | |
| wiki_bitcoin | | (+)* | | | | | | | | | | | | | | | | (-)* | |
| wiki_ethereum | | (+)** | | | | | | | | | | | | | | | | | |
| wiki_ripple | | | | | | | | | | | | | | | | | | | |
| wiki_litecoin | | | | (+)* | | | | | | | | | | | | | | | |
| wiki_monero | | | | | | | | | | | | | | | | | | | |
| wiki_dash | | | | | | | | | | | | | | | | | | | |
| wiki_nem | | | | | | | | | | | | | | | | | | | |
| wiki_dogecoin | | | | | | | | | | | | | | | | | | | |
| wiki_peercoin | | | | | | | | | (-)** | | | | | | | | | | |
| wiki_namecoin | | | | | | | | | | (+)* | | | | | | | | | |
| wiki_altcoins | (+)** | (+)*** | | | | | | | (+)* | (+)** | | (-)*** | | | | | | | (-)* |
| *Dummy variables* | | | | | | | | | | | | | | | | | | | |
| DUbitcoin | (-)*** | | | | | | | | | | | | | | | | | | |
| DUethereum | | (+)*** | | | | | | | | | | | | | | | | | |
| DUlitecoin | | | | (-)*** | | | | | | | | | | | | | | | |
| DUpeercoin | | | | | | | | | | | | | | | | | | | |
| DUnamecoin | | | | | | | | | | (+)** | | | | | | | | | |
| DUsupernet | | | | | | | | | | | | | | | | (-)** | | | |
| constant | (-)2*** | | | | | | | | | | | (-)*** | | | | | | | |
| trend | | | | | | | | | (+)*** | (+)*** | | | (+)* | | | | | | |

Notes: Dependent variable: prices of virtual currencies (including Altcoin indices). DU: dummy variable; *** significant at 1% level, ** significant at 5% level, * significant at 10% level. Empty cell indicates either absence of a variable in the respective model or the coefficient is not significantly different from zero. The sigh in parentheses means the sign of the estimated coefficient followed by the significance level.

**Table 7. Estimation results: short-run relationships between virtual currency prices**

| | BTC | ETH | XRP | LTC | XMR | DASH | NEM | DOGE | PPC | NMC | NVC | NXT | XCP | MINT | QORA | UNITY | BTS | alt19 | alt100usd |
|---|---|---|---|---|---|---|---|---|---|---|---|---|---|---|---|---|---|---|---|
| *Prices* | | | | | | | | | | | | | | | | | | | |
| bitcoin_usd | | (±)2* | (±)2** | (±)4** | (±)3* | (+)1*** | | (+)1** | (+)1* | (+)1* | (±)3* | (±)3* | (+)2** | (+)1* | (+)1* | (+)1** | (+)2** | (+)3*** | (-)1*** | (±)3** |
| alt19 | (-)1*** | (+)2* | (+)2** | (+)2* | (+)1* | | (+)2* | (+)3* | (+)3** | (+)3* | (±)3* | (+)1* | | (±)3** | (+)1*** | (+)2** | (+)1* | (-)1** | (+)2* |
| alt100usd | | | | | | | | | | | | | | | | | | (+)2** | (+)1** |
| Altcoin own price lags[1] | | (+)1* | (±)3* | (±)3* | (±)2* | (-)1* | (-)2* | (-)2* | (±)4* | (±)3* | (±)2* | (+)1* | | (-)4* | (-)1** | (-)2* | (±)3** | | |
| *Macro-financial variables* | | | | | | | | | | | | | | | | | | | |
| gold_price | (±)2** | | (-)1*** | | | | (-)1** | | (-)1* | (±)2** | (-)1** | (-)1** | | | | (-)1** | | | (+)1** |
| nasdaq | (+)1*** | | (+)2*** | | | | (-)1** | | | (+)1*** | | | | | | | | | |
| treasury_rate10y | (+)1*** | (+)1* | (-)1** | | | | (-)1*** | | (+)1** | (+)1** | | | | (+)1** | | | | | (+)1*** |
| e_usd_eur | | (±)3** | (+)1*** | (+)1*** | | (+)1*** | (±)2** | (-)1*** | | | (+)1** | (+)1*** | (+)1*** | | | | | (+)1*** | (-)2** |
| e_yuan_usd | (+)1** | (-)1*** | (-)1*** | | | | | | | | | | | | | (-)1*** | | | (-)1* |
| oil_price | | (-)2** | (-)1*** | | | | (+)1*** | | (±)2** | | | (+)1** | (-)1*** | (±)2*** | | (+)1* | | | (-)1* |
| *Wikipedia views* | | | | | | | | | | | | | | | | | | | |
| wiki_bitcoin | (-)1*** | (+)1** | (±)2* | | | | (-)1*** | | | (±)3* | | | | | | | | | |
| wiki_ethereum | | (+)2* | | | | | | | | | | | | | | | | | |
| wiki_ripple | | | (+)3* | | | | | | | | | | | | | | | | |
| wiki_litecoin | | | | (+)2* | | | | | | | | | | | | | | | |
| wiki_monero | | | | | (+)1* | | | | | | | | | | | | | | |
| wiki_dash | | | | | | (+)1* | | | | | | | | | | | | | |
| wiki_nem | | | | | | | (+)3** | | | | | | | | | | | | |
| wiki_dogecoin | | | | | | | | (+)2** | | | | | | | | | | | |
| wiki_peercoin | | | | | | | | | (+)2* | | | | | | | | | | |
| wiki_namecoin | | | | | | | | | | (+)6* | | | | | | | | | |
| wiki_altcoins | (-)1*** | (±)3** | (-)1*** | | | | (-)1*** | | (-)4** | (-)4** | (-)2* | (-)1* | (-)1*** | (+)2** | | | (+)1*** | | (±)2* |
| *Dummy variables* | | | | | | | | | | | | | | | | | | | |
| DUbitcoin | (-)1* | | | | | | | | | | | | | | | | | | |
| DUethereum | | (+)1* | | | | | | | | | | | | | | | | | |
| DUlitecoin | | | | (-)1* | | | | | | | | | | | | | | | |
| DUpeercoin | | | | | | | | | | | | | | | | | | | |
| DUnamecoin | | | | | | | | | | | (+)1** | | | | | | | | |
| DUsupernet | | | | | | | | | | | | | | | | (-)1** | | | |
| DUbitshares | | | | | | | | | | | | | | | | | (-)1** | | |
| *Supply of virtual currencies* | | | | | | | | | | | | | | | | | | | |
| supply_bitcoin | | | | | | | | | | | | (+)1** | | | | | | | |
| Altcoin own supply[2] | | (-)1* | | | (+)1** | | (-)1* | | | | | | | | | | | | |
| supply_altcoins | | | | | | | | | | | | (-)1*** | | | | | | | |
| Trend | | | (+)1* | | | | | | | | | (-)1*** | (+)1* | | (+)1** | (+1)** | | (+)1* | (+)1* |
| constant | | | | | | | | | (-)1* | | | | (-)1** | | | | (+)1* | (-)1** | (-)1** |

Notes: Dependent variable: prices of virtual currencies (including Altcoin indices). DU: dummy variable; *** significant at 1% level, ** significant at 5% level, * significant at 10% level. Empty cells indicate either absence of a variable in the respective model or the coefficient is not significantly different from zero. The sign in parentheses means the sign of the estimated coefficient followed by the number of significant lags (including the contemporary) and their significance level. [1]This row includes the significance level of virtual currencies' prices with respect to own price lags of ETH, XRP, LTC, XMR, DASH, NEM, DOGE, PPC, NMC, NVC, NXT, XCP, MINT, QORA, UNITY, BTS, respectively. [2]This row includes significance level of virtual currencies' prices with respect to own coin supply of ETH, XRP, LTC, XMR, DASH, NEM, DOGE, PPC, NMC, NVC, NXT, XCP, MINT, QORA, UNITY, BTS, respectively.

**Figure 1. Market capitalization of selected virtual currencies (billion $USD)**

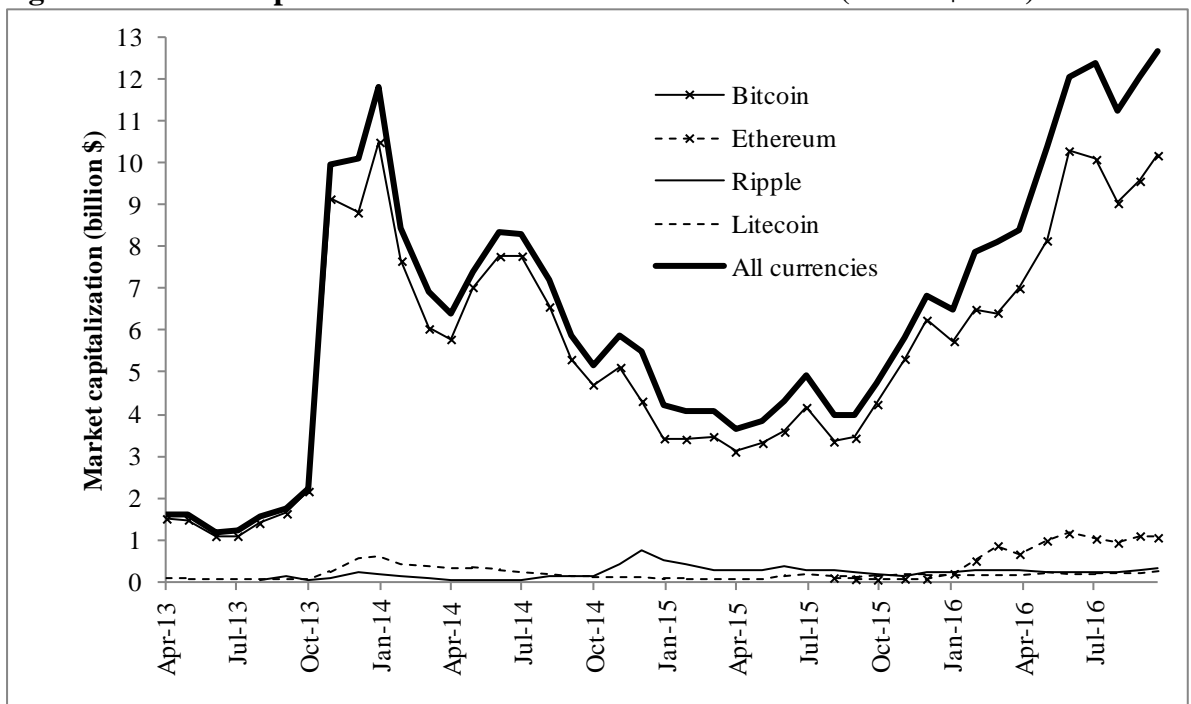

Source: coinmarketcap.com

**Figure 2. Market share of selected virtual currencies (%)**

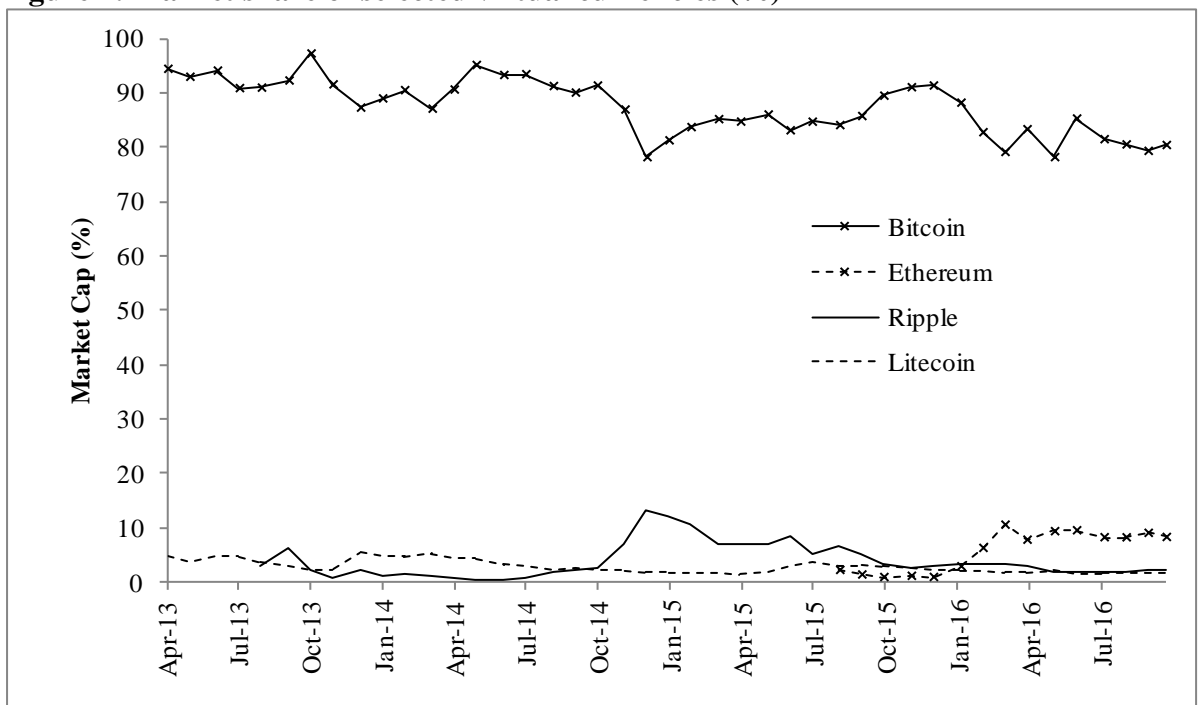

Source: coinmarketcap.com

**Figure 3. Price development of BitCoin and altcoins**

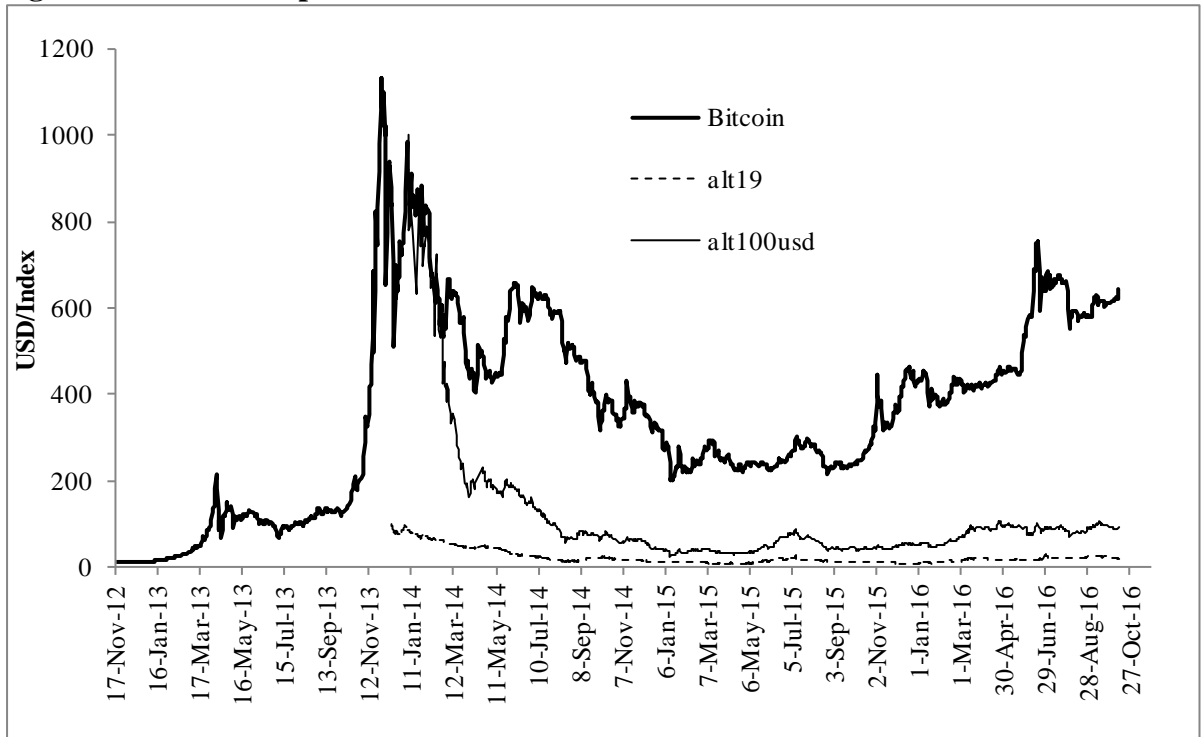

Source: BitCoin: quandl.com; alt19, alt100usd: alt19.com
Notes: BitCoin price: $USD per unit; alt19: index value calculated in BitCoin with base value of 100 set for 13 December 2013; alt100usd: index value is calculated in USD with base value of 1000 set for 6 January 2014.

**Figure 4. Currency composition of the global BitCoin trading volume (%)**

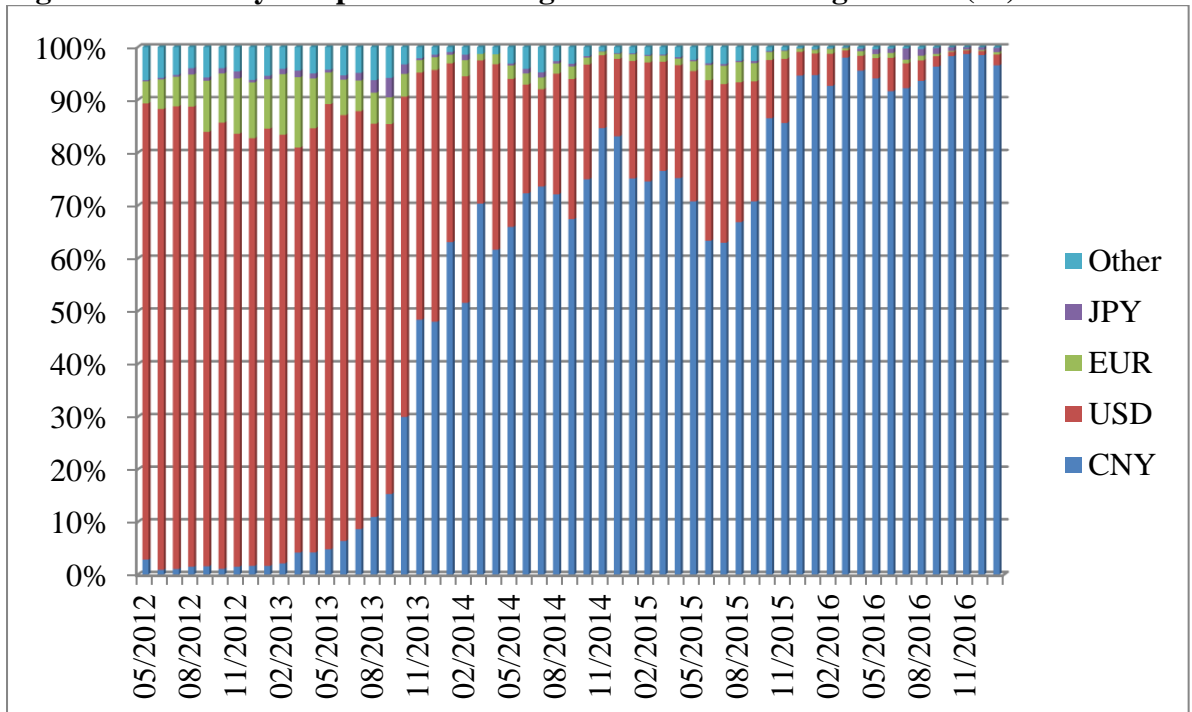

Source: http://data.BitCoinity.org/

**Figure 5. Coin supply development for virtual currencies with maximum supply limit**

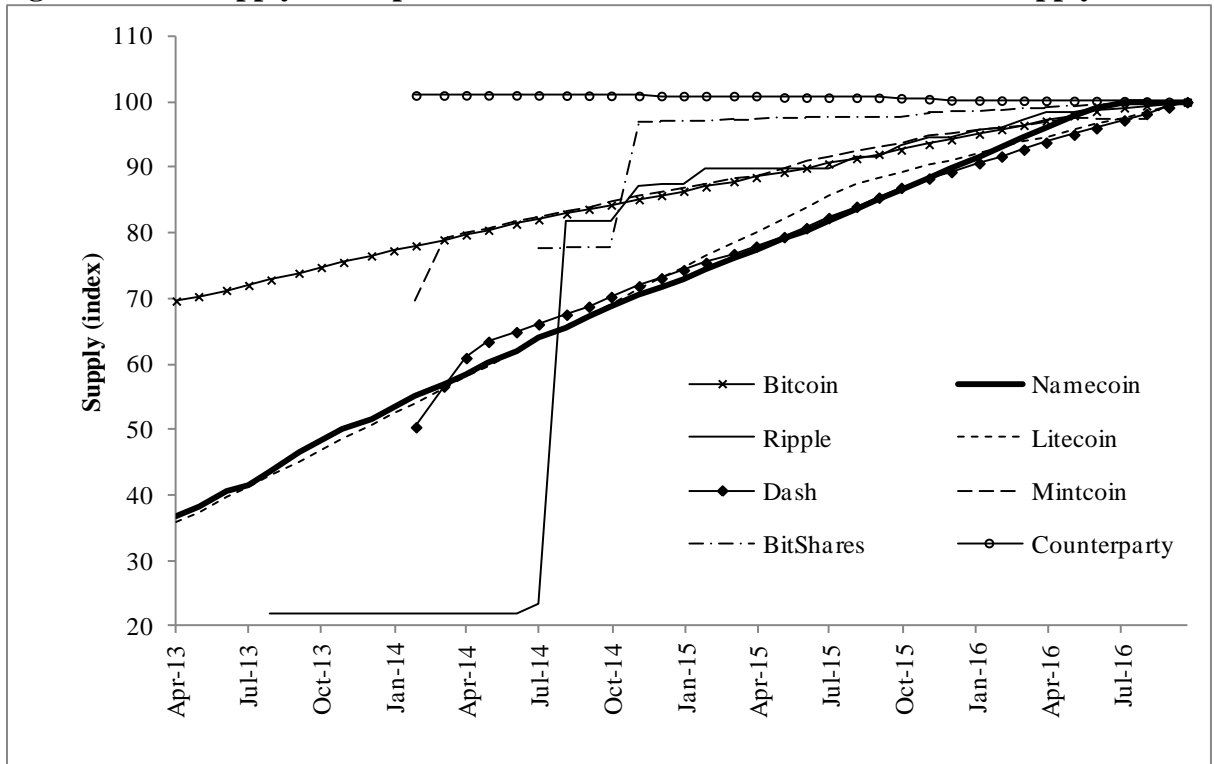

Source: coinmarketcap.com

**Figure 6. Coin supply development for virtual currencies with unlimited supply**

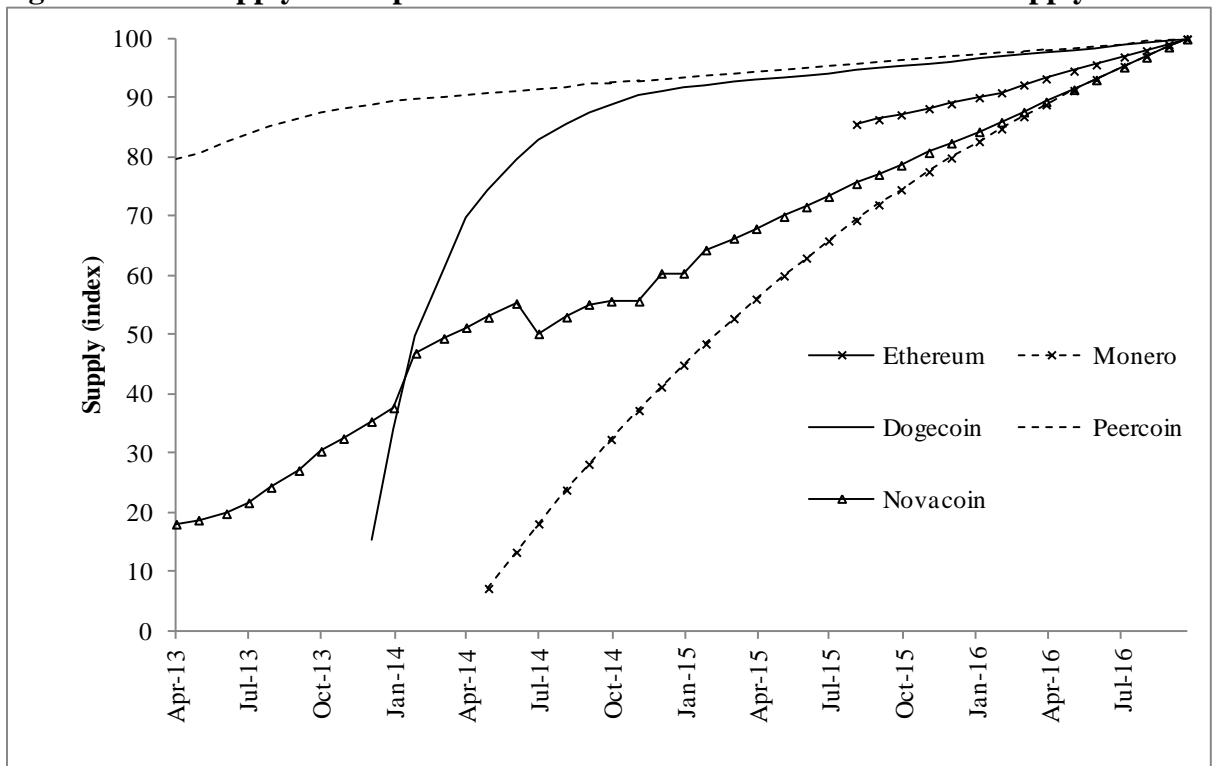

Source: coinmarketcap.com

**Appendix 1: Distributed ledgers and Blockchain**

An underlying characteristic of a virtual currency is the distributed ledger (i.e. a decentralized payment system), which is a publicly available mechanism used to verify and accept transactions (block) and shared among all network participants. Being a fully decentralized payment system, there are no intermediaries (e.g. banks) or a central authority (e.g. central bank) that would control the virtual currency. The decentralization is possible, because every network participant can check any transaction on the ledger, which serves as a proof of all past transactions (Ali et al. 2014).

The ledger of virtual currencies is built on a technology called blockchain, which allows transactions to be securely stored and verified without a centralized authority. Blockchain keeps a register of ownership by recording all transactions and their time chronology. The main principle of blockchain is that a set of transactions are recorded on separate blocks and blocks are linked in a sequential order without the possibility to change the information recorded on them retroactively. Almost all altcoins rely to the same underlying technology as BitCoin – blockchain. However, different modifications/improvements have been introduced to the canonical transaction verification mechanism of BitCoin.

The creation of ledger requires undertaking a costly action. That is, changing the ledger (recording transactions in the blockchain) requires a deployment of resources by network participants (miners). These resources include electricity, the computing power, time, temporary giving-up coins and represents the cost to secure and maintain the virtual currency network. The aim of undertaking this costly action is to deter abuses of the virtual currency. Miners willing to deploy their resources incur these and other costs which in return ensures a securely maintained the virtual currency network. In return, they are compensated for their costs by receiving transaction fees and/or newly minted coins (Ali et al. 2014; Farell 2015).